%% file: main.tex
\pdfoutput=1

\documentclass[acmtog]{acmart}
\acmSubmissionID{1743}
\usepackage{diagbox}
\usepackage{verbatim}
\usepackage{amssymb}
\usepackage{latexsym}
\usepackage{lineno}
\usepackage{xcolor}
\usepackage{tabularx}
\usepackage{xspace}
\usepackage{color}
\usepackage{stfloats}
\graphicspath{{figures/}}
\usepackage{amsmath,bm}
\usepackage{psfrag}
\usepackage{mathtools}
\usepackage{hyperref}

\usepackage{multirow, tabularx}
\newcolumntype{Y}{>{\centering\arraybackslash}X}
\usepackage{capt-of}%
\usepackage{array, boldline, makecell, booktabs}

\usepackage{colortbl}

\usepackage{array}
\usepackage{siunitx}

\usepackage{pifont}
\newcommand{\cmark}{\textcolor{green!50!black}{\ding{51}}}%
\newcommand{\xmark}{\textcolor{red!60!black}{\ding{55}}}%

\usepackage{psfrag}
\usepackage{forest}
\usepackage{makecell}
\usepackage{physics}
\usepackage{algorithm}
\usepackage{threeparttable}
\usepackage{graphicx}
\usepackage{subcaption}
\usepackage{flushend}
\usepackage{wrapfig}
\usepackage{algpseudocode}
\usepackage{fancyhdr}
\usepackage{algpseudocode}

\usepackage{url}
\usepackage{xcolor}
\definecolor{newcolor}{rgb}{.8,.349,.1}

\usepackage{enumitem}
\usepackage{svg}
\svgpath{{./img/}} 

\usepackage[normalem]{ulem}
\newcommand{\rev}[1]{#1}
\newcommand{\revv}[1]{}
\newcommand{\revvv}[1]{}

\begin{document}
\author{Sinan Wang}
\authornote{Both authors contributed equally to this research.}
\email{swang3081@gatech.edu}
\affiliation{
\institution{Georgia Institute of Technology}
\country{USA}
}

\author{Ruicheng Wang}
\authornotemark[1]
\email{wrc0326@outlook.com}
\affiliation{
\institution{Georgia Institute of Technology}
\country{USA}
}

\author{Taiyuan Zhang}
\email{imaginer.tai@gmail.com}
\affiliation{
\institution{Dartmouth College}
\country{USA}
}

\author{Fan Feng}
\email{fan.feng.gr@dartmouth.edu}
\affiliation{
\institution{Dartmouth College}
\country{USA}
}

\author{Jinjin He}
\email{jhe433@gatech.edu}
\affiliation{
\institution{Georgia Institute of Technology}
\country{USA}
}

\author{Yuchen Sun}
\email{yuchen.sun.eecs@gmail.com}
\affiliation{
\institution{Georgia Institute of Technology}
\country{USA}
}

\author{Zhiqi Li}
\email{zli3167@gatech.edu}
\affiliation{
\institution{Georgia Institute of Technology}
\country{USA}
}

\author{Bo Zhu}
\email{bo.zhu@gatech.edu}
\affiliation{
\institution{Georgia Institute of Technology}
\country{USA}
}

\title{Hamiltonian Two-Way Coupling of Nonlinear Waves and 3D Flows}
\begin{abstract}
Simulating large-scale free-surface \rev{flow} by coupling a localized 3D fluid solver to a cheaper 2D surface model has long \rev{faced a mismatch in wave dynamics: efficient 2D wave models used in graphics are typically either linear or non-dispersive. These models are fast, simple, and accurate for calm, small-amplitude seas, but coupling them to strongly nonlinear 3D solvers produces} visible reflections and artifacts at the 2D--3D interface. We address this problem by introducing a \revv{fully} nonlinear \revv{Hamiltonian} \rev{and dispersive} 2D wave model based on the canonical Zakharov formulation. Its Hamiltonian structure enables canonically consistent two-way coupling, allowing information to pass smoothly across the \rev{2D--3D interface}. Our 2D solver reduces mean wave-height error by \rev{ factors of $1.7$--$5\times$ relative to}  SWE, BEM, and Airy baselines while running more than $10^3\times$ faster than BEM\rev{; it achieves greater nonlinear accuracy and coupling fidelity than SWE and Airy, with minor losses in speed and stability}. \rev{Coupling it with a 3D Navier--Stokes solver yields a full system that suppresses visible seam artifacts across a range of experiments, including} dispersion-matching and Kelvin-wake tests, and runs over $4\times$ faster than a \rev{full-domain} GPU NB-FLIP simulation. 
\end{abstract}

\keywords{Fluid simulation, wave simulation}

\ccsdesc[500]{Computing methodologies~Physical simulation}

\begin{teaserfigure}
\centering
\includegraphics[width=1.0\textwidth]{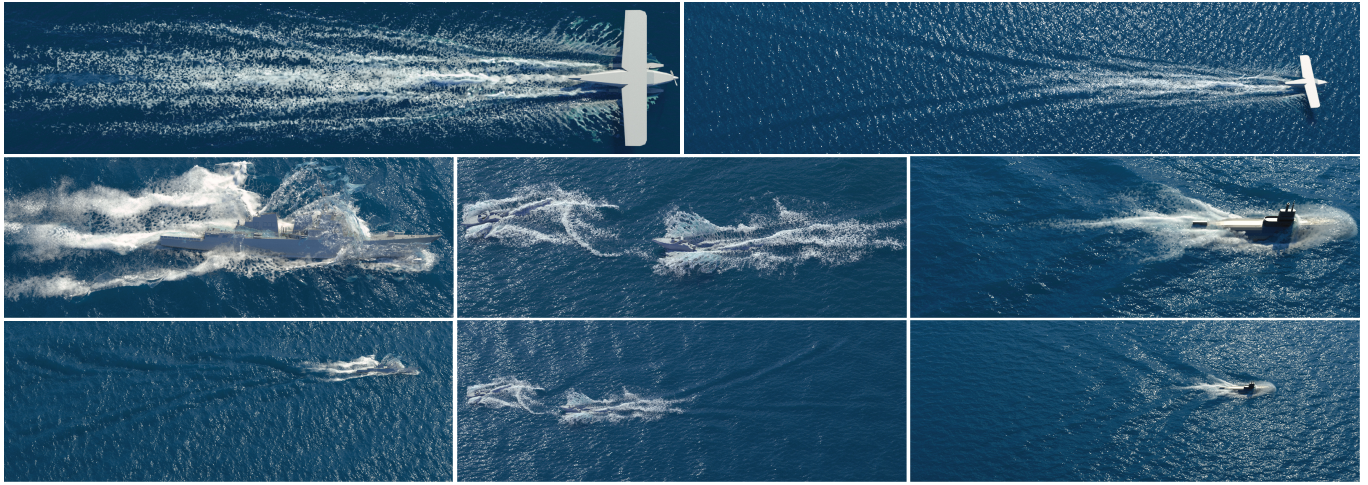}
\caption{A seaplane landing on the ocean surface, shown in near view (top-left) and far view (top-right). \rev{Near} views of a battleship navigating through heavy waves (middle-left), two boats chasing each other (middle-center), and a submarine rising to the ocean surface (middle-right), with their corresponding \rev{far} views shown in the bottom row.}
\label{fig:teaser}
\end{teaserfigure}

\setcopyright{cc}
\setcctype{by}
\acmJournal{TOG}
\acmYear{2026} \acmVolume{45} \acmNumber{6} \acmArticle{179}
\acmMonth{12} \acmDOI{10.1145/3842540}

\maketitle

\input{intro}
\input{related_work}
\input{wave}

\input{2Dto3D}
\input{3Dto2D}
\input{implementation}

\input{time_int}
\input{validation}
\input{conclusion_limitation}


{
\bibliography{refs_ML_sim.bib, refs_INR.bib, refs_flow_map.bib, refs_simulation.bib, refs_vortex_method, refs_2D, refs_couple}
\bibliographystyle{ACM-Reference-Format}
}

\end{document}

%% file: intro.tex
\section{Introduction}
\label{sec:intro}

Simulating large-scale free-surface \rev{flow} remains a long-standing computational bottleneck in graphics, whether the scene is a seaplane splashing down on the open ocean, a battleship plowing through heavy seas, or a submarine breaking the ocean surface. A fully three-dimensional incompressible Navier--Stokes (NS) simulation captures every regime of interest, including splashes, breaking waves, and fluid--solid coupling, but its memory and compute costs grow prohibitive once the domain spans hundreds of meters. A practical remedy is to confine the expensive 3D solver to a small embedding region and model the surrounding open ocean with a much cheaper 2D surface theory \cite{thurey2006animation, chentanez2015coupling,huang2021ships, schreck2022coupling,stomakhin2017fluxed} in order to propagate far-field waves efficiently. The success of this 2D--3D decomposition rests on the quality of the 2D model and the fidelity of the coupling between the two solvers.

\rev{Within this 2D--3D decomposition, a key obstacle to practical, high-fidelity coupling is that graphics currently lacks an efficient 2D wave model that is simultaneously \emph{nonlinear and dispersive}: none retains the wave--wave interactions whose strength grows with wave steepness (nonlinearity) while allowing waves of different wavelengths to travel at their correct, wavelength-dependent speeds (dispersion)}. The two wave theories that dominate the literature both bypass the nonlinear surface dynamics. Airy wave theory \cite{airy1845tides, tessendorf2014ewave,canabal2016dispersion, schreck2019fundamental} linearizes the free-surface boundary conditions around the flat state, retaining the correct dispersion $\omega^2 = gk\tanh(kh)$ but discarding every nonlinear interaction that gives rise to scenarios like the steepening of Stokes wavetrains. \rev{The discarded terms scale with wave steepness, so for the gentle ambient swell that Airy solvers typically animate, the linearization costs little and buys great speed and simplicity.} The shallow-water equations (SWE) \cite{kass1990rapid, chentanez2010real, layton2002numerically} carry nonlinear depth-averaged advection but collapse the dispersion relation to $\omega = \sqrt{gh}\,k$, removing the very dispersive structure that distinguishes a real ocean from a sloshing tank\rev{, a sacrifice that is acceptable in the transport-dominated shallow flows SWE was built for, but not in open water}. Lagrangian alternatives that have gained traction in graphics, e.g., wave particles~\cite{yuksel2007wave}, wave packets~\cite{jeschke2017water}, wavefront tracking~\cite{jeschke2015water}, and water-surface wavelets~\cite{jeschke2018water}, all transport their primitives along linear ray paths and inherit the same linearization. \rev{Each of these trade-offs is defensible in its intended regime. Once strongly nonlinear waves dominate, however, these models lose accuracy, whereas our method makes the opposite trade-off, sacrificing a little speed and stability in exchange for greater nonlinear accuracy and coupling fidelity.}

Coupling a localized 3D NS solver to such a linear 2D \rev{wave model} \rev{creates precisely such a strongly nonlinear setting}. \rev{The 3D solver is strongly nonlinear by construction, so waves reaching the embedding-box boundary exhibit nonlinear effects such as crest steepening and amplitude-dependent dispersion.} A linear 2D model cannot represent any of these, so the two \rev{solvers} disagree on the very content of the surface state at the interface. The disagreement shows up as spurious reflections at the seam and visible surface artifacts. We therefore identify the absence of a \revv{fully} nonlinear\revv{, Hamiltonian} and dispersive 2D wave model in graphics as \rev{a key obstacle to a 2D--3D coupling pipeline that is at once practical, accurate, and largely free of interface artifacts}. The one prior nonlinear route, the boundary-element method (BEM)~\cite{huang2021ships}, recovers the \revv{fully} nonlinear free-surface potential flow at the PDE level, but discretizes it on a Lagrangian surface mesh, and pays for its nonlinear \rev{accuracy} with per-step surface tracking/remeshing, dense $\mathcal{O}(N^3)$ boundary interactions, \revv{a non-GPU-native framework,} and high implementation complexity.

\rev{We address this limitation by introducing a 2D Eulerian wave model that is at once \emph{nonlinear} and \emph{dispersive}}\rev{, and we show that its underlying \emph{Hamiltonian} structure is what makes the two-way coupling to a 3D NS solver natural}. The \revv{2D side}\rev{wave model} evolves the canonical Zakharov pair \cite{zakharov1968stability}: the surface elevation $\eta$ and the surface value $\psi$ of the velocity potential. \rev{Here \emph{Hamiltonian} means that $(\eta,\psi)$ form a canonical pair in the sense of classical mechanics: they evolve like position and momentum under the total wave energy, the Hamiltonian (Section~\ref{sec:wave}).} \rev{We advance the Zakharov system using the Craig--Sulem expansion of the Dirichlet--Neumann operator (DNO) $G(\eta)$ \citep{craig1993numerical}. This reduces every nonlinear correction to} FFTs and pointwise products on a regular grid at $\mathcal{O}(N^2\log N)$ cost. \rev{In practice, we retain this expansion through second or third order (\autoref{sec:wave}).} \rev{The same DNO machinery enables two-way coupling: it lifts $(\eta,\psi)$ to depth-resolved boundary velocities for the 3D solver and, in reverse, reconstitutes a canonical wave state after relaxing $\eta$ toward the 3D surface.}



On standalone 2D wave-propagation benchmarks, our wave model alone lowers the mean wave-height error by \revv{roughly $2\times$}\rev{$1.7\times$ (deep water) to $2.4\times$ (shallow water)} over the strongest linear baseline (Airy with dispersion kernels) and by $3$ to $5\times$ over SWE and BEM, while running more than $10^3\times$ faster than BEM\rev{; the price of this accuracy is a modestly more expensive and slightly less stable wave step than the linear baselines, a few milliseconds per substep in absolute terms (\autoref{tab:wave_prop_time_comp})}. Once coupled to the 3D NS solver, a dispersion-matching test in which a wave packet radiates from the 3D box across the seam and a single-boat experiment in which a hull radiates a Kelvin wake through the seam together show that our coupled pipeline yields symmetric\revv{, artifact-free} wave fields \rev{with minimal seam artifacts}, whereas \rev{prior coupling schemes exhibit varying degrees of asymmetric propagation, spurious reflection, or visible 2D--3D inconsistency (\autoref{sec:validation})}; on the single-boat scene, our solver also runs more than $4\times$ faster than a GPU NB-FLIP simulation of the same domain\rev{, i.e., the bounding box of the entire visible water region}. We further run a suite of large-scale scenarios, a battleship in heavy seas, a surfacing submarine, two \rev{chasing boats}, and a seaplane landing on the open ocean, in which the 2D--3D interface remains \rev{largely free of visible artifacts}.
\paragraph{Contributions} Our main contributions are:
\begin{enumerate}[leftmargin=*,labelindent=0pt,itemindent=0pt,listparindent=0pt]
\item \textbf{A \revv{Hamiltonian, fully} nonlinear and dispersive Eulerian wave model for graphics.} We evolve the canonical Zakharov pair $(\eta,\psi)$ under the exact water-wave Hamiltonian on a structured grid at FFT cost, capturing \revv{fully} nonlinear surface phenomena \rev{ such as the steepening of Stokes wavetrains}. \rev{This retains the nonlinear modeling capability} of the prior BEM \rev{approach while reducing computational complexity from} $\mathcal{O}(N^3)$ to $\mathcal{O}(N^2\log N)$ without requiring per-step surface tracking or remeshing.

\item \textbf{A canonically consistent two-way coupling between $(\eta,\psi)$ and a 3D NS solver.} Both directions \rev{build on the DNO expansion}: the forward direction harmonically extends $(\eta,\psi)$ to depth-resolved bulk velocities on the embedding-box faces, while the reverse direction relaxes the \rev{wave model's} $\eta$ toward the \rev{3D free-surface elevation} and reconstitutes $\psi$ from the relaxed $\eta$ and the wave's own surface velocity through an inverse DNO solve, keeping the wave state canonical at every substep.

\item \textbf{\rev{A continuously tunable nonlinear wave model}.} A scalar perturbation parameter $\varepsilon\in[0,1]$ modulates every nonlinear correction in the truncated DNO and the Bernoulli equation \rev{for stability}, parametrizing a continuous family of solvers that degenerates to linear Airy theory at $\varepsilon=0$ and recovers the \revv{fully} nonlinear \rev{(truncated)} Zakharov system at $\varepsilon=1$.
\end{enumerate}

%% file: related_work.tex
\section{Related Work}
\subsection{2D simulation}


\paragraph{SWE}
The SWE, a simplified form of the Navier--Stokes equations, originated in engineering physics \cite{le2013introduction, crapper1984introduction, stoker2019water} and were first introduced to computer graphics by \citet{kass1990rapid}. \citet{layton2002numerically} introduced an implicit semi-Lagrangian scheme to improve stability and support larger time steps. \citet{chentanez2010real} extended SWE to handle arbitrary slopes and depths. Later works enhanced SWE by adding surface tension \cite{wang2007solving}, reducing numerical dissipation while ensuring stability via implicit Newmark integration \cite{angst2008robust}, and improving efficiency using a collocated grid and a simplified SWE formulation \cite{lee2007fast}.

\paragraph{Airy Wave Theory}
Airy wave theory \cite{airy1845tides, dean1991water, birkhoff2015hydrodynamics}, popularized in computer graphics by \citet{tessendorf2001simulating}, synthesizes ocean height fields by evolving wave spectra according to dispersion \rev{relations}. \rev{Some implementations augment linearly evolved waves with geometric displacement or post-sharpening \cite{tessendorf2001simulating,stomakhin2017fluxed}; we distinguish this geometric nonlinearity from the dynamical nonlinearity in our wave evolution.} \citet{horvath2015empirical} further improved spectral realism using an advanced empirical model, and proposed an artist-friendly parameterization scheme to facilitate intuitive control over ocean wave elongation. Beyond Tessendorf's formulation, \citet{starn2001simple} proposed a compact FFT-based fluid solver widely used in 2D surface simulations.
\citet{nielsen2013synthesizing} fit artist-created wave fields to spectral parameters for consistent high-frequency synthesis.
Later, \citet{tessendorf2014ewave} presented the \textit{eWave} model, which provides a time integration in both Fourier and convolution forms, enabling more flexible boundary treatments. Dispersion-kernel approaches \cite{canabal2016dispersion} handle solid boundaries via convolution and shadow masks. More recently, \citet{schreck2019fundamental} introduced a fundamental solution-based method using Green's functions to capture wave--obstacle interactions in open domains without grids.

\paragraph{High-Order Spectral Method}
Starting from the canonical Hamiltonian formulation of \citet{zakharov1968stability}, the high-order spectral (HOS) method was independently introduced by \citet{dommermuth1987high} and \citet{west1987new} to evolve \revv{fully} nonlinear surface waves on a flat reference grid via a Taylor expansion of the Dirichlet--Neumann operator, with \citet{craig1993numerical} casting the expansion in the explicit operator form used in modern HOS solvers. \rev{Such interior-to-boundary reductions via Dirichlet--Neumann (Steklov--Poincar\'e) operators also appear in graphics, including extrinsic shape analysis \cite{wang2018steklov}, boundary-first conformal flattening \cite{sawhney2017boundary}, and boundary-only physics-based skinning \cite{gao2014steklov}.}
Subsequent work extended HOS to directional sea states \cite{bateman2001efficient}, variable bathymetry \cite{guyenne2008high}, and large-scale open-ocean simulation \cite{ducrozet2016hos}. We adopt the HOS expansion as the engine of our 2D wave solver.

\paragraph{Other Wave Simulation}
A variety of approaches have explored alternative physical representations to better capture dispersion, boundary interactions, and visual detail. These include wave particles \cite{yuksel2007wave}, wave packets \cite{jeschke2017water}, wavefront tracking \cite{jeschke2015water}, and water surface wavelets \cite{jeschke2018water}. Hybrid models combine different regimes, such as SWE with Airy theory \cite{jeschke2023generalizing}, or augment height fields with particles for breaking waves \cite{thurey2007real}. Others enhance detail as a post-process, like wave curves on deforming surfaces \cite{skrivan2020wave}.

\subsection{3D simulation}
Early works in 3D fluid simulation for graphics \cite{stam1999stable, foster1996realistic, foster2001practical} primarily adopt Eulerian frameworks, but often suffer from numerical dissipation. Pure Lagrangian approaches, such as smoothed particle hydrodynamics (SPH) \cite{muller2003particle, ihmsen2014sph}, require costly neighbor searches and struggle to enforce incompressibility. Hybrid Lagrangian-Eulerian methods such as Particle-in-Cell (PIC) \cite{harlow1964particle} alleviate some issues but still exhibit dissipation. To address this, the Fluid-Implicit Particle (FLIP) method was introduced to graphics by \citet{zhu2005animating}, building on \citet{brackbill1988flip}, significantly reducing dissipation. To improve efficiency, Narrow Band FLIP (NB-FLIP) \cite{ferstl2016narrow} and extended NB-FLIP \cite{sato2018extended} were later proposed.
Furthermore, several approaches \cite{losasso2004simulating, losasso2006spatially, zhu2013new, nielsen2016spatially, ando2020practical, wang2025cirrus} have employed adaptive techniques to resolve fine-scale fluid features while maintaining computational efficiency for large-scale scenes.
Recently, flow-map-based methods \cite{nabizadeh2022covector, deng2023fluid,  zhou2024eulerian, chen2024solid, chen2025neural, chen2025fluid, he2026level, wang2024eulerian, wang2026two, wang2025fluid, li2025clebsch, li2025edge, li2024particle, li2026impulse, sun2024impulse, sun2025lfm} and the Coadjoint Orbit FLIP \cite{nabizadeh2024coflip} have reduced numerical dissipation and better preserved vortices.

\begin{table}[t]
\centering
\caption{Comparison with prior 2D--3D coupled water-simulation methods. \rev{``GPU-native'' means that the 2D solver advances a height field stored on a uniform structured grid, making it well suited for GPU implementation; }``Mesh-op-free'' means the 2D solver requires no per-step mesh operations on an unstructured surface mesh (mesh construction, tracking, or remeshing) during simulation, in contrast to BEM-style methods that maintain a Lagrangian surface mesh; ``Warmup-free'' means the 2D solver is usable from $t=0$ without first accumulating a temporal sampling window; \rev{``Nonlinear'' means nonlinear wave dynamics rather than geometric sharpening; }``Dispersive'' means the 2D wave model reproduces the correct deep-water dispersion relation, so waves of different wavelengths propagate at different speeds\rev{; ``Hamiltonian'' means the 2D solver evolves the canonical Zakharov pair $(\eta,\psi)$ under the water-wave Hamiltonian (Section~\ref{sec:wave})}.}
\label{tab:related_compare}
\resizebox{\columnwidth}{!}{%
\begin{tabular}{lccccccc}
\toprule
Method & Hamiltonian & Two-way & Nonlinear & GPU-native & Mesh-op-free & Warmup-free & Dispersive \\
\midrule
\citet{thurey2006animation}                 & \xmark & \cmark & \cmark & \cmark & \cmark & \cmark & \xmark \\
FAB~\cite{stomakhin2017fluxed}              & \xmark & \xmark & \xmark & \cmark & \cmark & \cmark & \cmark \\
\citet{chentanez2015coupling}               & \xmark & \cmark & \cmark & \cmark & \cmark & \cmark & \xmark \\
\citet{schreck2022coupling}                 & \xmark & \cmark & \xmark & \cmark & \cmark & \xmark & \cmark \\
\citet{huang2021ships}                      & \xmark & \cmark & \cmark & \xmark & \xmark & \cmark & \cmark \\
\textbf{Ours}                               & \cmark & \cmark & \cmark & \cmark & \cmark & \cmark & \cmark \\
\bottomrule
\end{tabular}%
}
\end{table}

\subsection{Coupled 2D and 3D simulation} \citet{thurey2006animation} \rev{coupled} shallow water with a 3D solver using the Lattice Boltzmann method. \citet{nielsen2011guide} improved art-directability by introducing guide shapes to constrain high-resolution liquid simulations to low-resolution inputs.
\citet{chentanez2015coupling} combined SWE, SPH, and a 3D Eulerian solver, enabling efficient simulation but lacking accurate wave dispersion. \citet{huang2021ships} addressed this by coupling BEM \cite{da2016surface, huang2020surface} with FLIP for more realistic ocean–ship interaction, though BEM's mesh operations are computationally expensive and limit scalability. The fluxed animated boundary (FAB) method \cite{stomakhin2017fluxed} allows artistic control but only supports one-way coupling (the 3D simulation cannot affect the 2D simulation). Perfectly Matched Layers (PML)-based methods \cite{berenger1994perfectly, bojsen2016generalized, soderstrom2010pml} reduce reflection at the interface via non-reflecting boundaries and allow localized re-simulation, but are also one-way. More recently, \citet{schreck2022coupling} coupled a 2D fundamental solution solver \cite{schreck2019fundamental} with a 3D FLIP simulation, using dispersion matching to reduce coupling artifacts. Nonetheless, \rev{its underlying wave model remains linear Airy theory and therefore does not capture nonlinear wave effects}.
Table~\ref{tab:related_compare} summarizes how our method compares to representative prior coupled 2D--3D water solvers. To our knowledge, ours is the first to combine \revv{fully} nonlinear and dispersive surface-wave dynamics with two-way coupling\revv{, GPU-native structured-grid evaluation,} and ocean-scale efficiency in a single framework.

%% file: wave.tex
\section{Hamiltonian Spectral Wave Solver}
\label{sec:wave}

We model the open-ocean portion of the domain as the free surface of an incompressible, irrotational, inviscid fluid of finite depth $-d \le y \le \eta(x,z,t)$ under gravity $g$. Throughout this section we let $\bm{x}=(x,z)$ denote the horizontal coordinates, $y$ the vertical coordinate, $\eta(\bm{x},t)$ the surface elevation, $\Phi(\bm{x},y,t)$ the bulk velocity potential ($\bm u = \nabla\Phi$, $\Delta\Phi=0$), and $\psi(\bm{x},t) = \Phi(\bm{x},\eta(\bm{x},t),t)$ the velocity potential evaluated on the moving free surface. \rev{Our full method couples the 2D wave solver developed in this section to a localized 3D NS solver (Sections~\ref{sec:2d_to_3d} and~\ref{sec:3d_to_2d}); to keep the two components clearly distinguished, we refer to the former as the \emph{wave} model and to the latter as the \emph{NS} solver; when we speak of the ``2D (wave) side'' and the ``3D (NS) side'', we mean these two subdomains and their respective solvers. The dimension labels describe our default configuration; the same construction applies verbatim one dimension lower, where a 1D wave model embeds a 2D NS solver, as in some of our validation experiments (\autoref{sec:validation}).} Symbol conventions are summarized in Table~\ref{tab:notation_table}.

\newcolumntype{z}{X}
\newcolumntype{s}{>{\hsize=.25\hsize}X}
\begin{table}[!t]
\caption{Summary of the main symbols and notations.}
\centering
\small
\begin{tabularx}{0.47\textwidth}{scz}
\hlineB{2.5}
Notation & Type & Definition\\
\hlineB{2.5}
\hspace{12pt}$\eta$ & scalar & free-surface elevation\\
\hline
\hspace{12pt}$\psi$ & scalar & surface velocity potential $\Phi|_{\mathrm{surface}}$\\
\hline
\hspace{12pt}$\Phi$ & scalar & bulk velocity potential, $\bm u = \nabla\Phi$\\
\hline
\hspace{12pt}$G(\eta)$ & operator & Dirichlet--Neumann operator (DNO)\\
\hline
\hspace{12pt}$G_0$ & operator & DNO at the flat state $\eta\equiv 0$\\
\hline
\hspace{12pt}$B_0$ & operator & Fourier multiplier with symbol $|\bm{k}|^2$\\
\hline
\hspace{12pt}$\bm{k}$ & vector & horizontal wave vector\\
\hline
\hspace{12pt}$\omega$ & scalar & wave angular frequency, $\omega^2 = g\,G_0$\\
\hline
\hspace{12pt}$\bm u$ & vector & velocity, components $(u, v, w)$\\
\hline
\hspace{12pt}$\varepsilon$ & scalar & nonlinear-amplitude parameter\\
\hline
\hspace{12pt}$d$ & scalar & still-water depth\\
\hline
\hspace{12pt}$\sigma(\bm x)$ & scalar & relaxation source \rev{rate}\\
\hline
\hspace{12pt}$\phi$ & scalar & 3D level-set field\\
\hline
\hspace{12pt}$t$ & scalar & time\\
\hlineB{2.5}
\end{tabularx}
\captionsetup{aboveskip=20pt}
\label{tab:notation_table}
\end{table}

\subsection{Canonical Zakharov formulation}
\label{sec:zakharov}

Following \citet{zakharov1968stability}, the water-wave problem admits a Hamiltonian formulation in which the canonical pair is exactly $(\eta,\psi)$. The fluid Hamiltonian is the sum of kinetic and potential energy, expressed entirely in surface variables,
\begin{equation}
H[\eta,\psi] \;=\; \tfrac{1}{2}\!\int \psi\,G(\eta)\psi \,d\bm{x} \;+\; \tfrac{1}{2}g\!\int \eta^2\,d\bm{x},
\label{eq:hamiltonian}
\end{equation}
and the equations of motion are
\begin{equation}
\partial_t\eta \;=\; \frac{\delta H}{\delta\psi}, \qquad \partial_t\psi \;=\; -\frac{\delta H}{\delta\eta}.
\label{eq:hamilton_eom}
\end{equation}
The single nontrivial object in this formulation is the Dirichlet--Neumann operator (DNO) $G(\eta)$, defined by
\begin{equation}
G(\eta)\psi(\bm{x}) \;=\; \sqrt{1+|\nabla\eta|^2}\;\partial_n\Phi\big|_{y=\eta},
\label{eq:dno_def}
\end{equation}
i.e.\rev{,} the operator that, for each prescribed Dirichlet datum $\psi$ on the moving surface, returns the normal derivative of the harmonic extension $\Phi$ into the fluid. Knowing $G(\eta)$ collapses the original three-dimensional Laplace problem to two surface fields $(\eta,\psi)$ on a flat horizontal grid.

Carrying out the variational derivatives in \eqref{eq:hamilton_eom} yields the canonical Zakharov system (\citet{zakharov1968stability}, recast by \citet{craig1993numerical}),
\begin{align}
\partial_t \eta &= G(\eta)\,\psi,
\label{eq:zakharov_kin}\\[2pt]
\partial_t \psi &= -g\eta - \tfrac{1}{2}|\nabla\psi|^2 + \frac{\bigl(G(\eta)\psi + \nabla\eta\cdot\nabla\psi\bigr)^2}{2\bigl(1+|\nabla\eta|^2\bigr)}.
\label{eq:zakharov_dyn}
\end{align}
Equation~\eqref{eq:zakharov_kin} is the kinematic free-surface boundary condition expressed through the DNO; \eqref{eq:zakharov_dyn} is the exact Bernoulli equation for the surface potential. The system is closed: a numerical method that can evaluate $G(\eta)\psi$ for arbitrary $(\eta,\psi)$ on a regular grid yields a \revv{fully} nonlinear ocean solver. \rev{Throughout the paper, ``Hamiltonian'' and ``canonical'' refer to this Zakharov state $(\eta,\psi)$ and its energy structure. This structure is not a prerequisite for high-quality coupling in general; its value here is that $(\eta,\psi)$ completely describes the irrotational surface state, so both directions of the 2D--3D exchange reduce to well-defined maps on these two fields: the 2D-to-3D direction (Section~\ref{sec:2d_to_3d}) lifts $(\eta,\psi)$ to a depth-resolved bulk velocity, and the 3D-to-2D direction (Section~\ref{sec:3d_to_2d}) folds the 3D surface state back into a valid pair.}

\begin{figure}[!t]
    \centering
    \includegraphics[width=0.5\textwidth]{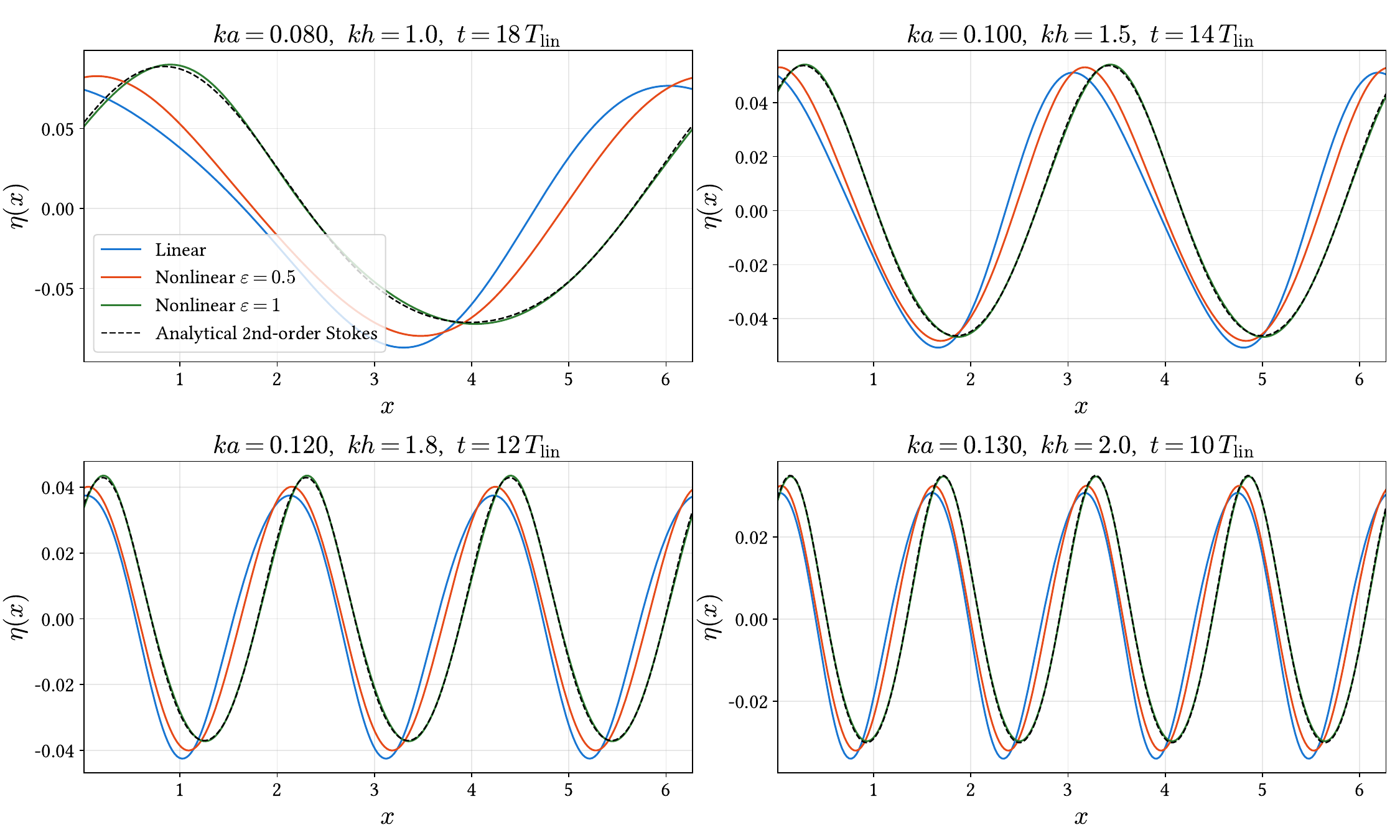}
    \caption{Final-time surface elevation $\eta(x)$ of the linear wave solver ($\varepsilon=0$, blue, equivalent to \cite{canabal2016dispersion}), our nonlinear High-Order Spectral (HOS)-3 solver \rev{with $\varepsilon=0.5$ (red) and $\varepsilon=1$ (green)}, and the analytical second-order Stokes wave (black dashed) in four mild Stokes regimes\rev{, i.e., weakly nonlinear, non-breaking Stokes waves}. The linear solver drifts out of phase due to the missing amplitude-dependent frequency shift, while the \rev{fully nonlinear ($\varepsilon=1$)} HOS-3 solver stays phase-locked to the analytical wave. \rev{The $\varepsilon=0.5$ curve lies between the two.}}
    \label{fig:illustration}
\end{figure}

\subsection{Spectral expansion of the Dirichlet--Neumann operator}
\label{sec:dno_expansion}

A direct evaluation of $G(\eta)\psi$ requires solving Laplace's equation in the irregular fluid domain at every step. \citet{craig1993numerical} observed that, for $\eta$ small relative to the characteristic wavelength $1/k$, $G(\eta)$ admits a convergent Taylor expansion in homogeneous powers of $\eta$ that involves no Laplace solve at all. We adopt this expansion, known in ocean engineering as the High-Order Spectral (HOS) method~\cite{dommermuth1987high}, as the engine of our wave solver.
Let $G_0$ denote the operator at the flat reference state $\eta\equiv 0$. Its symbol on a horizontal Fourier mode $e^{i\bm{k}\cdot\bm{x}}$ is
\begin{equation}
G_0(\bm{k}) \;=\; |\bm{k}|\tanh(|\bm{k}|d),
\label{eq:G0}
\end{equation}
which is exactly the linear water-wave dispersion operator and reproduces $\omega^2(\bm k) = g\,G_0(\bm k)$ at leading order \cite{airy1845tides}. Writing $G(\eta) = \sum_{j\ge 0} G_j(\eta)$ with $G_j$ homogeneous of degree $j$ in $\eta$, the recursion of \citet{craig1993numerical} gives, in spectral form,
\begin{align}
G_1(\eta)\psi &= -G_0\bigl(\eta\,G_0\psi\bigr) \;-\; \nabla\!\cdot\!\bigl(\eta\,\nabla\psi\bigr),
\label{eq:G1}\\
G_2(\eta)\psi &=\;\; G_0\bigl(\eta\,G_0(\eta\,G_0\psi)\bigr) - \tfrac{1}{2}G_0\bigl(\eta^2\,B_0\psi\bigr) \nonumber\\
&\quad -\eta\,B_0(\eta\,G_0\psi) + \tfrac{1}{2}\eta^2\,G_0(B_0\psi) + |\nabla\eta|^2\,G_0\psi,
\label{eq:G2}
\end{align}
where $B_0$ is the homogeneous Fourier multiplier with symbol $|\bm k|^2$ (the second vertical derivative of the harmonic extension at the flat state). All operators in \eqref{eq:G1}--\eqref{eq:G2} are concatenations of Fourier multipliers ($G_0$, $B_0$, $\nabla$) with pointwise products of $\eta$, $\nabla\eta$, $\nabla\psi$, and are therefore evaluated with FFTs and physical-space multiplications.


\rev{To improve stability in coupled simulations with steep 3D-driven surface features,} we modulate every nonlinear term by a scalar amplitude parameter $\varepsilon\in[0,1]$,
\begin{equation}
G(\eta) \;\approx\; G_0 \;+\; \varepsilon\,G_1(\eta) \;+\; \varepsilon^2 G_2(\eta),
\label{eq:dno_truncated}
\end{equation}
and read $\varepsilon=1$ as the fully nonlinear model and $\varepsilon=0$ as the linear Airy limit. The dynamic equation \eqref{eq:zakharov_dyn} is rewritten consistently as
\begin{equation}
\partial_t\psi \;=\; -g\eta + \varepsilon\,\Bigl[-\tfrac{1}{2}|\nabla\psi|^2 + \tfrac{1}{2}\bigl(1+\varepsilon^2|\nabla\eta|^2\bigr)\,W^2\Bigr],
\label{eq:dyn_truncated}
\end{equation}
with
\begin{equation}
W \;=\; \frac{G_0\psi + \varepsilon\,G_1(\eta)\psi + \varepsilon^2\,G_2(\eta)\psi + \varepsilon\,\nabla\eta\cdot\nabla\psi}{1+\varepsilon^2|\nabla\eta|^2}.
\label{eq:W}
\end{equation}

Throughout the paper we refer to the truncation in \eqref{eq:dno_truncated}, which retains the three terms $G_0$, $\varepsilon\,G_1$, $\varepsilon^2\,G_2$, as the Order-3 (or HOS-3) scheme; the Order-2 (HOS-2) variant referenced in our experiments retains only $G_0$ and $\varepsilon\,G_1$.

\newlength{\wtlabelraise}\setlength{\wtlabelraise}{2.0em}
\begin{figure*}[!htbp]
    \centering
    \begin{subfigure}[b]{0.3333\linewidth}
        \centering
        \makebox[0pt][r]{\raisebox{\wtlabelraise}{\rotatebox{90}{\footnotesize Linear}}\hspace{1pt}}%
        \includegraphics[width=\linewidth]{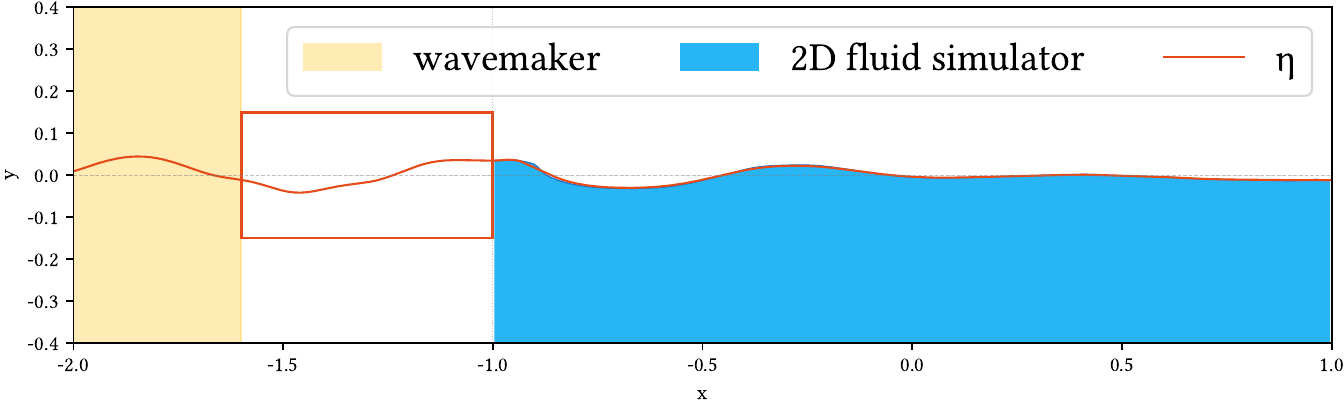}
        \makebox[0pt][r]{\raisebox{\wtlabelraise}{\rotatebox{90}{\footnotesize Nonlinear}}\hspace{1pt}}%
        \includegraphics[width=\linewidth]{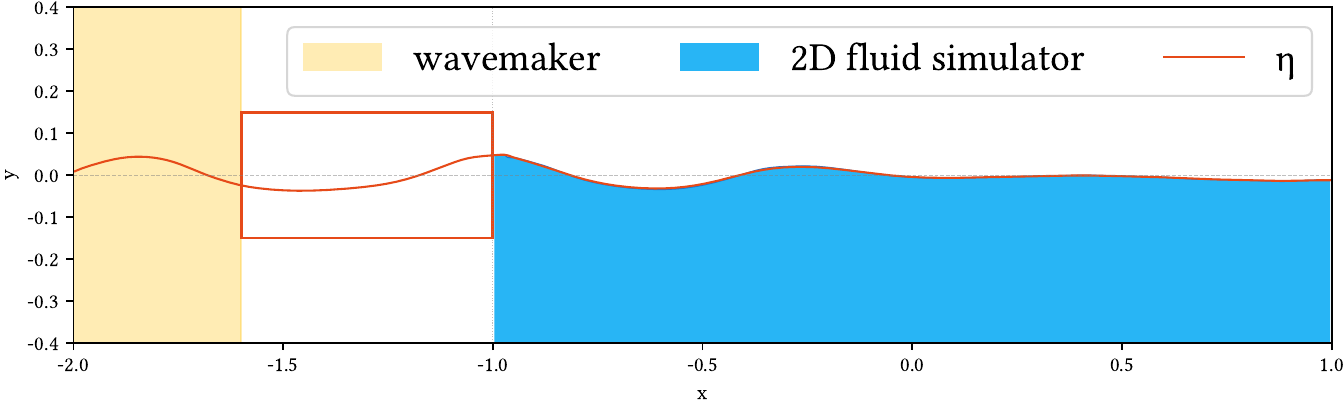}
        \caption{$t\approx\SI{2.65}{\second}$ ($\approx 3.7$ periods)}
        \label{fig:wavetrain_t1}
    \end{subfigure}%
    \begin{subfigure}[b]{0.3333\linewidth}
        \centering
        \includegraphics[width=\linewidth]{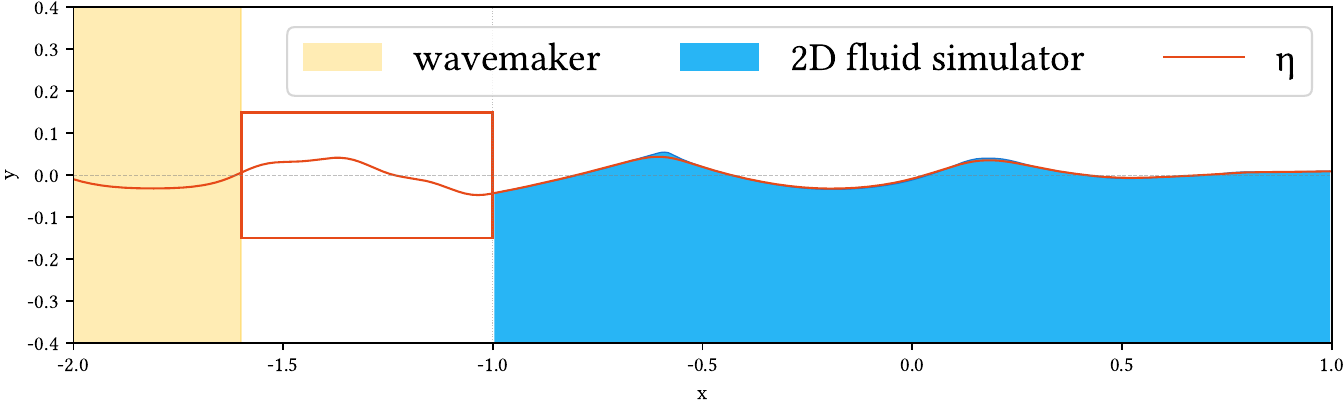}
        \includegraphics[width=\linewidth]{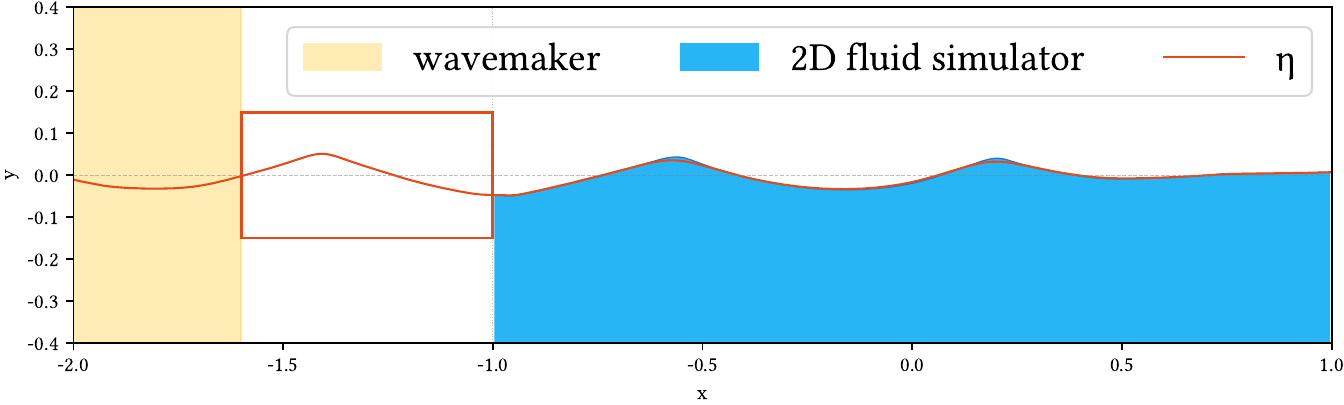}
        \caption{$t\approx\SI{3.75}{\second}$ ($\approx 5.2$ periods)}
        \label{fig:wavetrain_t2}
    \end{subfigure}%
    \begin{subfigure}[b]{0.3333\linewidth}
        \centering
        \includegraphics[width=\linewidth]{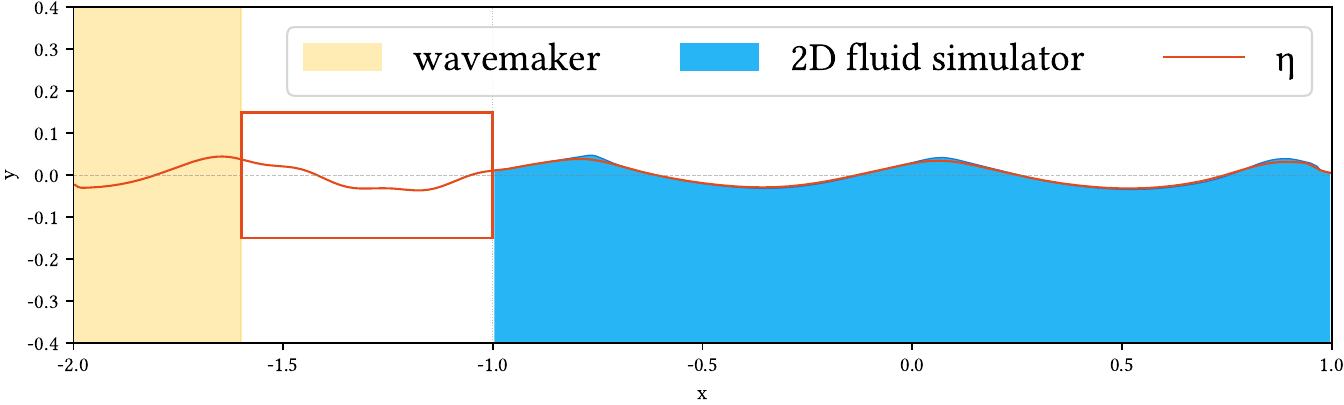}
        \includegraphics[width=\linewidth]{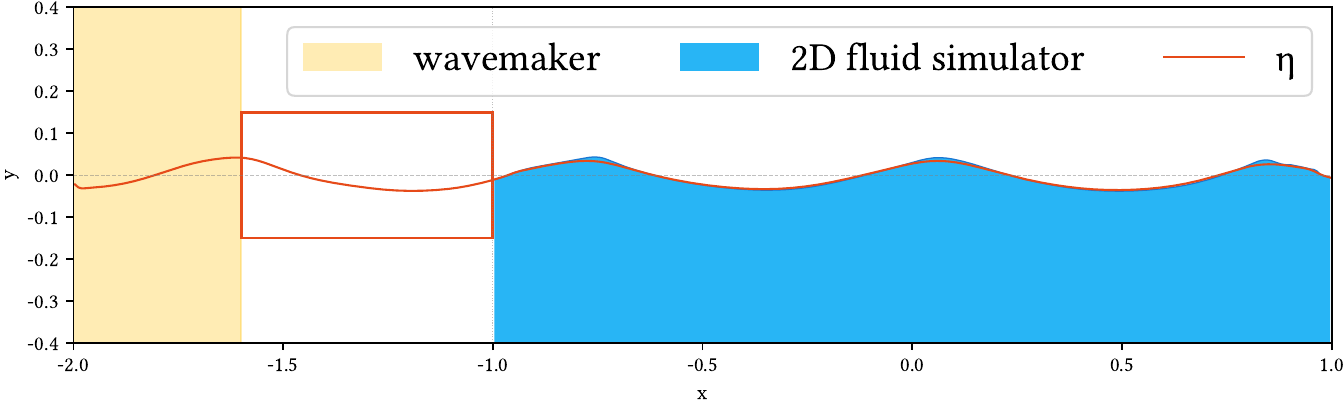}
        \caption{$t\approx\SI{5.0}{\second}$ ($\approx 7.0$ periods)}
        \label{fig:wavetrain_t3}
    \end{subfigure}
    \caption{Continuous Stokes wavetrain crossing the 2D--3D coupling interface, comparing the linear wave model ($\varepsilon=0$, top row of each column) with our fully nonlinear HOS model ($\varepsilon=1$, bottom row). A 1D wave domain ($x\in[-2,\,1]\,\si{\meter}$, still-water depth \SI{0.4}{\meter}) drives a second-order Stokes train of steepness $ka=0.3$ and wavelength \SI{0.8}{\meter} through an embedded 2D fluid box ($x\in[-1,\,1]\,\si{\meter}$, light blue fill); yellow shading marks the left-edge wavemaker zone\revv{ and grey shading marks the right-side sponge}. The red boxes outline the wave free-propagation region just outside the fluid box, where the 1D wave solver carries the train on its own. Inside the red boxes the linear surface develops visible kinks, is no longer smooth, and shows no crest--trough asymmetry, while the nonlinear surface remains smooth, exhibits the sharpened crests and flattened troughs of a Stokes train, and continues the surface shape inside the 2D fluid box almost exactly. The mismatch in the linear case reflects the incompatibility between the Airy wave model and the natively nonlinear 2D \rev{NS} solver across the coupling interface \rev{ at this steepness ($ka=0.3$)}.}
    \label{fig:wavetrain}
\end{figure*}

We verify the effectiveness of this $\varepsilon$ modulation in Section~\ref{sec:validation}: the wave-propagation sweep against a high-resolution 3D Eulerian reference (\autoref{fig:wave_prop_err}) and the Stokes-wave sweep against the analytical second-order Stokes solution (\autoref{fig:eps_sweep_err}) both show the height error decreasing monotonically as $\varepsilon$ grows from $0$ to $1$, with the drop between $\varepsilon=0$ and $\varepsilon\approx 0.1$ already accounting for a large fraction of the linear-model error. In other words, even a small $\varepsilon$ already yields a significant improvement over the linear case. Conversely, in scenarios where strong forcing pushes the surface close to the breaking threshold and the higher-order DNO terms can excite numerical instabilities, we deliberately keep $\varepsilon$ small to preserve solver stability: $\varepsilon=0.3$ for the submarine, $\varepsilon=0.2$ for the battleship in heavy waves, and $\varepsilon=0.2$ for the pond.

\subsection{Boundary conditions via even extension}
\label{sec:bc}

The FFT machinery above is intrinsically periodic; closed boundaries (a tank wall, a coastline, the perimeter of a pond) require additional care. We support reflecting and replicating boundaries through an even extension of the wave domain. Given the physical state $(\eta,\psi)$ on a rectangle of resolution $(M,N)$, we form
\begin{equation}
\tilde{\eta}(\bm{x}) \;=\; \eta\bigl(R\bm{x}\bigr), \qquad \tilde\psi(\bm{x}) \;=\; \psi\bigl(R\bm{x}\bigr),
\label{eq:even_ext}
\end{equation}
on a doubled $(2M,2N)$ grid, where $R$ denotes the appropriate reflection across each boundary. The operators $G_0$, $G_1$, and $G_2$ are then constructed on the doubled grid and applied to $\tilde\eta$, $\tilde\psi$; after the time step we crop the result back to $(M,N)$. The construction makes the discrete normal derivatives at the boundary vanish identically (Neumann condition) for reflecting walls, while a replicate variant approximates outflow. Because the extension \rev{only doubles the grid,} the FFT cost increases only by a constant factor of four, and the boundary treatment carries no asymptotic overhead.

\subsection{Time integration: \rev{exact-linear} Adams--Bashforth}
\label{sec:time_step}

\rev{Following \citet{craig1993numerical}, we split the system into an exactly integrated linear part and an AB2-advanced nonlinear part:}
\begin{equation}
\partial_t \begin{pmatrix}\eta\\\psi\end{pmatrix}
\;=\; \mathcal L\begin{pmatrix}\eta\\\psi\end{pmatrix} \;+\; \mathcal N(\eta,\psi),
\label{eq:split}
\end{equation}
where the linear part $\mathcal L = \bigl(\begin{smallmatrix} 0 & G_0 \\ -g & 0 \end{smallmatrix}\bigr)$ is diagonal in Fourier space with eigenfrequency $\omega(\bm k) = \sqrt{g\,G_0(\bm k)}$, and the nonlinear forcing $\mathcal N$ collects all terms in $\varepsilon$. The linear part is stiff at high wavenumber, i.e., $\omega(\bm k)$ scales like $\sqrt{|\bm k|}$ in deep water, so an explicit step on the full RHS would require $\Delta t \lesssim 1/\omega(k_{\mathrm{Nyq}})$.
We avoid the stiffness by integrating $\mathcal L$ exactly. Per Fourier mode, the linear flow is the rotation
\begin{equation}
\begin{pmatrix} \hat\eta\\ \hat\psi \end{pmatrix}_{\text{lin}}\!(t+\Delta t)
\;=\; \begin{pmatrix} \cos(\omega \Delta t) & \tfrac{G_0}{\omega}\sin(\omega \Delta t) \\[3pt] -\tfrac{g}{\omega}\sin(\omega\Delta t) & \cos(\omega\Delta t) \end{pmatrix}\!\begin{pmatrix} \hat\eta\\\hat\psi \end{pmatrix}\!(t),
\label{eq:lin_prop}
\end{equation}
with the limit $\hat\eta\,{\to}\,\hat\eta,\ \hat\psi\,{\to}\,\hat\psi - g\Delta t\,\hat\eta$ at $\bm k = \bm 0$. The nonlinear forcing $\hat{\mathcal N}_n = (\hat F^\eta_n, \hat F^\psi_n)$ is then advanced through the Duhamel response of \eqref{eq:lin_prop}\rev{, i.e., the variation-of-constants formula, which propagates the accumulated nonlinear forcing with the exact linear flow}: setting
$\hat F^{q,\star}_n
= \tfrac{3}{2}\hat F^q_n
- \tfrac{1}{2}\hat F^q_{n-1}$,
$q\in\{\eta,\psi\}$
(an Adams--Bashforth-2 extrapolation), we update
\begin{align}
\hat\eta_{n+1} &= \cos(\omega\Delta t)\,\hat\eta_n + \tfrac{G_0}{\omega}\sin(\omega\Delta t)\,\hat\psi_n \nonumber\\
&\quad +\;\frac{\sin(\omega\Delta t)}{\omega}\,\hat F^{\eta,\star}_n \;+\; \frac{1-\cos(\omega\Delta t)}{g}\,\hat F^{\psi,\star}_n,
\label{eq:if_ab2_eta}\\[2pt]
\hat\psi_{n+1} &= -\tfrac{g}{\omega}\sin(\omega\Delta t)\,\hat\eta_n + \cos(\omega\Delta t)\,\hat\psi_n \nonumber\\
&\quad -\;\frac{1-\cos(\omega\Delta t)}{G_0}\,\hat F^{\eta,\star}_n \;+\; \frac{\sin(\omega\Delta t)}{\omega}\,\hat F^{\psi,\star}_n.
\label{eq:if_ab2_psi}
\end{align}

\begin{figure}[!t]
    \centering
    \includegraphics[width=0.48\textwidth]{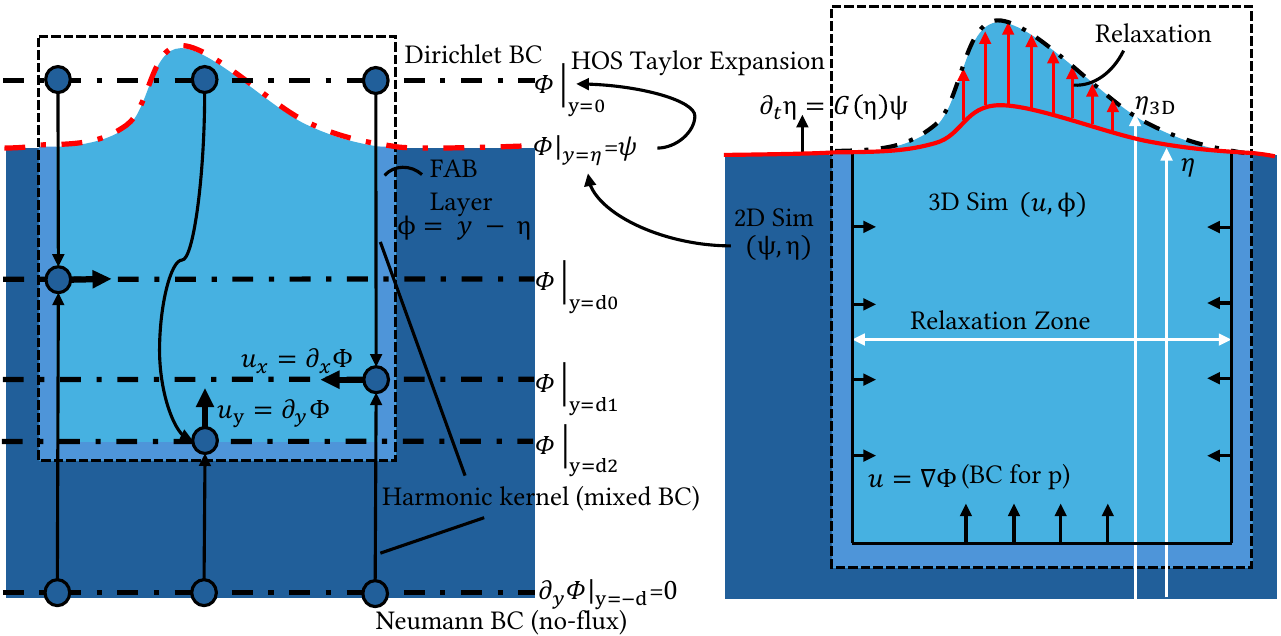}
    \caption{Two-way coupling between the 2D wave solver and the 3D NS solver. \textbf{Left, 2D-to-3D (Section~\ref{sec:2d_to_3d}):} $(\eta,\psi)$ is lifted to $y=0$ via the Craig--Sulem (HOS) Taylor expansion and propagated to depth layers by finite-depth harmonic kernels under Dirichlet top and Neumann no-flux bottom $\partial_y\Phi|_{y=-d}=0$. The FAB ring on the box faces takes its level-set BC from the 2D height field via $\phi=y-\eta$ and its velocity BC from $\bm u=\nabla\Phi$. \textbf{Right, 3D-to-2D (Section~\ref{sec:3d_to_2d}):} an exponential-decay source on the box footprint relaxes the wave-side $\eta$ toward the 3D surface $\eta_{3\mathrm D}$ implicitly inside the relaxation zone, while outside the box the Zakharov equation $\partial_t\eta=G(\eta)\psi$ evolves the ocean freely.}
    \label{fig:ocean_layers}
\end{figure}

The first step (and any step in which $\Delta t$ changes) bootstraps with $F^\star_0 = \mathcal N_0$, which reduces \eqref{eq:if_ab2_eta}--\eqref{eq:if_ab2_psi} to a first-order integrating-factor Euler step. The $k=0$ singularities cancel analytically: $\sin(\omega\Delta t)/\omega\to\Delta t$ and $(1-\cos(\omega\Delta t))/G_0 \to \tfrac{1}{2}g\,\Delta t^2$.

The scheme has three concrete advantages for our coupled setting. First, \eqref{eq:lin_prop} integrates the linear dispersion exactly per Fourier mode, so the surface step is non-dissipative on linear waves regardless of $\Delta t$. Second, the AB2 extrapolation only touches the nonlinear forcing, which is bounded; the time-step constraint reduces from $\omega(k_{\mathrm{Nyq}})\Delta t < 1$ to a much milder CFL on $\mathcal N$. Third, the same exponential-integrator strategy extends to the implicit relaxation source we add in Section~\ref{sec:3d_to_2d}: that source is linear in $\eta-\eta^{\mathrm{target}}$ with a position-dependent rate $\sigma(\bm x)$, so it admits its own closed-form exponential decay~\eqref{eq:eta_relax_exact}, which we compose with the wave step by operator splitting (Step~8 of Alg.~\ref{alg:main}). Each half is therefore integrated exactly over $\Delta t$, and the time-step constraint never tightens with $\sigma_{\max}$.
After each step we apply the spectral state filter of \citet{craig1993numerical},
\begin{equation}
A(\bm{k}) \;=\; \prod_{\alpha\in\{x,z\}} \tfrac{1}{8}\Bigl[5 + 4\cos\bigl(\pi |k_\alpha|/k_{\mathrm{Nyq}}^\alpha\bigr) - \cos\bigl(2\pi |k_\alpha|/k_{\mathrm{Nyq}}^\alpha\bigr)\Bigr],
\label{eq:state_filter}
\end{equation}
to $(\eta,\psi)$. The filter is unity at low wavenumbers, smoothly suppresses the highest modes, and removes residual high-frequency noise without affecting the resolved spectrum.

\subsection{State recovery: forward and inverse DNO maps}
\label{sec:state_recovery}

The wave solver evolves the canonical pair $(\eta,\psi)$, but the surrounding pipeline routinely needs to translate between this canonical state and the physically observable surface vertical velocity $v = G(\eta)\psi$. The \emph{forward} direction $(\eta,\psi)\mapsto v$ is the right-hand side of the kinematic Hamilton equation~\eqref{eq:zakharov_kin}, evaluated every substep by the time integrator. The \emph{inverse} direction $(\eta,v)\mapsto\psi$ appears in two places. First, in the 3D-to-2D relaxation of Section~\ref{sec:3d_to_2d}, the canonical conjugate $\psi$ is reconstituted from the post-relaxation pair $(\eta,v)$ to keep the wave state self-consistent across the relaxation step. Second, initial conditions and analytically prescribed sea states are naturally specified in physical $(\eta,v)$ form and must be lifted back to the canonical pair before the time integrator can advance them. We therefore expose both maps as utilities.

\begin{figure}[!t]
    \centering
    \includegraphics[width=0.48\textwidth]{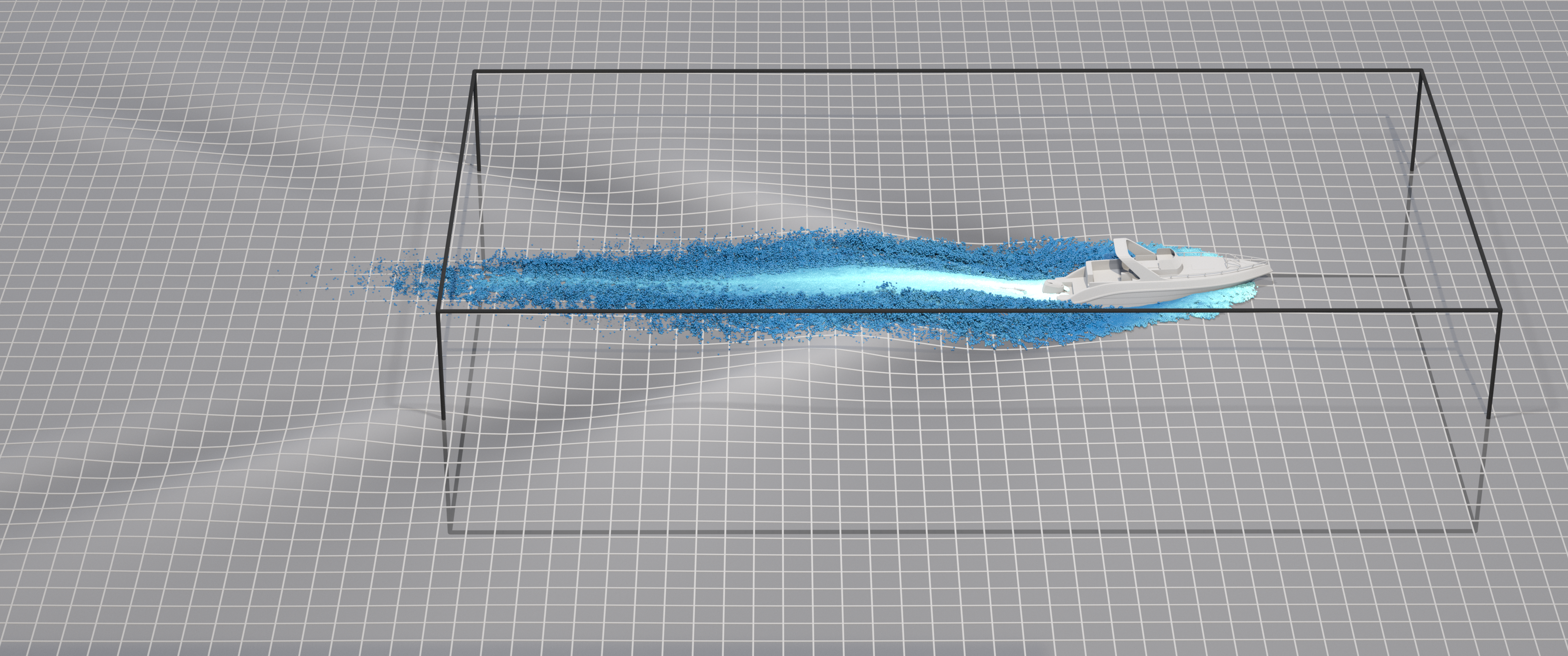}\hfill
    \includegraphics[width=0.48\textwidth]{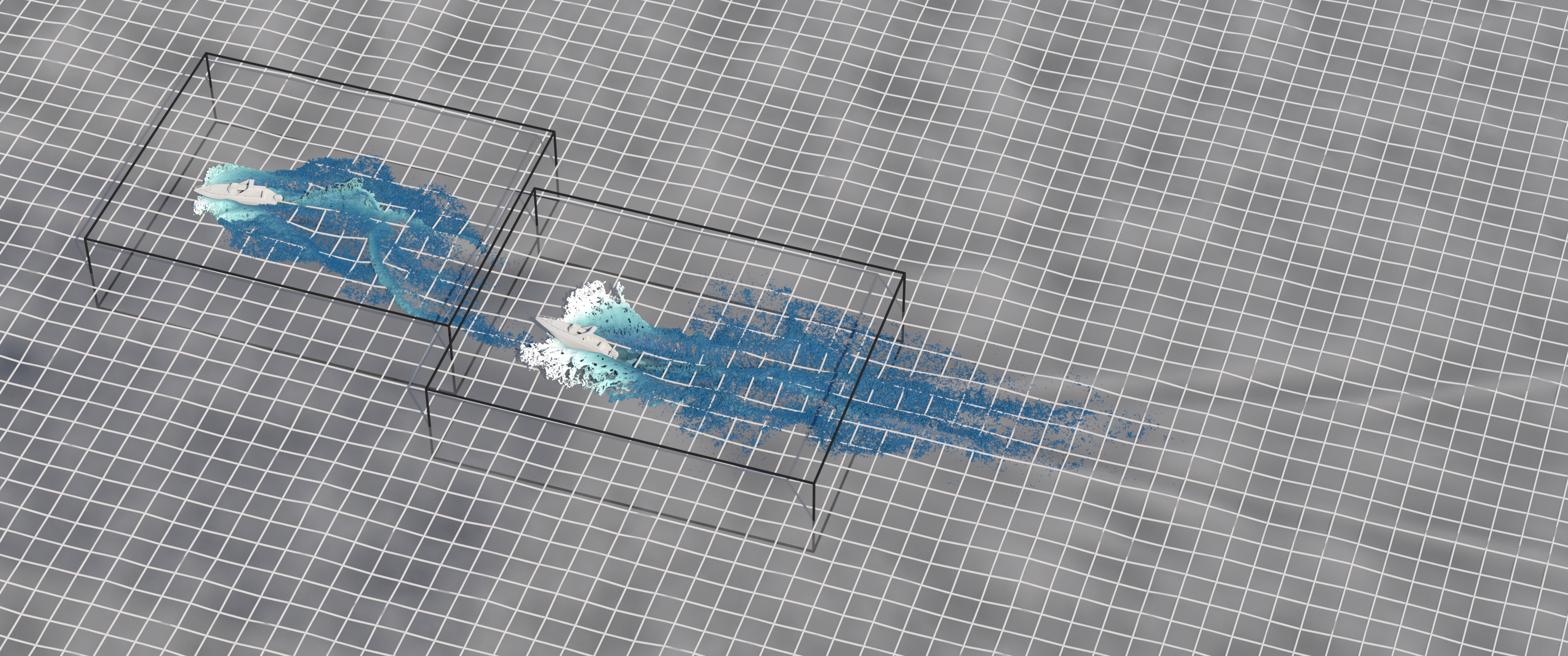}
    \caption{Overview of our two-way coupled method: a localized 3D NB-FLIP solver inside each embedding box (black outline) resolves the local 3D flow, and a 2D \revv{Hamiltonian} \rev{nonlinear} wave solver evolves the surrounding ocean on a structured grid. Multiple 3D regions can coexist and interact through the shared wave field (bottom).}
    \label{fig:method_overview}
\end{figure}

\paragraph{Forward map.}
The forward map is a single DNO application,
\begin{equation}
v \;=\; G_0\psi + \varepsilon\,G_1(\eta)\psi + \varepsilon^2\,G_2(\eta)\psi,
\label{eq:v_forward}
\end{equation}
evaluated by the same de-aliased FFT pipeline (Section~\ref{sec:dealias}).

\paragraph{Inverse map.}
The inverse map, needed whenever the wave state is initialized from a physical $(\eta,v)$ pair or when the relaxation source modifies $\eta$ relative to $\psi$, is obtained by fixed-point iteration on
\begin{equation}
\psi^{(n+1)} \;=\; G_0^{-1}\!\bigl(v - \varepsilon\,G_1(\eta)\psi^{(n)} - \varepsilon^2 G_2(\eta)\psi^{(n)}\bigr),
\label{eq:psi_inverse}
\end{equation}
seeded with $\psi^{(0)} = G_0^{-1}v$. Because $\varepsilon\,G_1$ and $\varepsilon^2\,G_2$ are subdominant relative to $G_0$ in the parameter regimes of interest, the iteration is contractive and converges in a handful of passes; we use eight iterations throughout. \rev{Each iteration evaluates the nonlinear DNO correction once and applies $G_0^{-1}$ through pointwise division in Fourier space, yielding $\mathcal{O}(N^2\log N)$ cost per iteration.}


%% file: 2Dto3D.tex
\section{2D-to-3D Coupling: Bulk Velocity from the Canonical Surface State}
\label{sec:2d_to_3d}

\begin{figure*}[!htbp]
    \centering
    \includegraphics[width=1.0\textwidth]{figures/dispersion_compare_fix0.8.pdf}
    \caption{Comparison of four 2D--3D coupling methods in the dispersion matching test, all driven by the same 3D NB-FLIP solver. The method of \citet{chentanez2015coupling} exhibits clear left--right asymmetry in the 2D wave field. The method of \rev{\citet{schreck2022coupling}, run with its dispersion-law rescaling factor recalibrated to our 3D solver configuration ($d=0.8$; see \autoref{sec:validation}), removes most of the left--right asymmetry, but its 2D and 3D regions remain less consistent than ours. FAB \cite{stomakhin2017fluxed} combined with the Airy Dispersion Kernel (DK) wave model \cite{canabal2016dispersion} shows no visible asymmetry, but produces visibly deeper waves inside the 3D region than in the surrounding 2D region.} Our method produces a wave field that is nearly indistinguishable across the 2D and 3D regions\rev{, while exhibiting smoother waveforms in the 3D region}. \rev{The wave-solver times are \SI{2.3}{\milli\second}, \SI{27.5}{\milli\second}, \SI{3.7}{\milli\second}, and \SI{13.3}{\milli\second} per substep for \citet{chentanez2015coupling}, \citet{schreck2022coupling}, FAB + Airy DK, and our method, respectively.} \rev{Our wave model uses $\varepsilon=1$.}}
    \label{fig:dispersion_compare}
\end{figure*}

The 3D NS solver lives in a localized embedding box $\Omega_{3\mathrm D}\subset\mathbb R^3$ surrounding the area of interest (a hull, a falling object, an outflow); see \autoref{fig:method_overview} for the overall layout of the coupled system. Its lateral and bottom faces sit \emph{inside} the open ocean simulated by the wave model of Section~\ref{sec:wave}, and at every substep the 3D solver requires a velocity boundary condition on the cells that lie outside the air phase but on the box perimeter. We build that boundary condition by reconstructing the bulk velocity field implied by the canonical surface state $(\eta,\psi)$ through finite depth, depth layer by depth layer; the left panel of \autoref{fig:ocean_layers} illustrates this pipeline, from the HOS lift of $(\eta,\psi)$ to $\Phi|_{y=0}$ through the harmonic continuation down to the depth layers from which $\bm u=\nabla\Phi$ is read off. The reverse half of the loop, in which the 3D surface state is folded back into the canonical pair through a relaxation source on the wave-side $\eta$ equation, is deferred to Section~\ref{sec:3d_to_2d}; in this section we treat the relaxation as a black box and focus solely on lifting $(\eta,\psi)$ to a depth-resolved bulk velocity on the FAB ring.

\subsection{The FAB layer: coupling boundary of the 3D box}
\label{sec:fab_layer}

Following the FAB method of \citet{stomakhin2017fluxed}, we mark the outermost band of cells on the lateral and bottom faces of $\Omega_{3\mathrm D}$ as a \emph{FAB} layer of width $b_{\mathrm{FAB}}$ (four cells in all our runs); the band is drawn in the left panel of \autoref{fig:ocean_layers}. FAB cells are excluded from the 3D pressure projection and instead carry Dirichlet data prescribed by the wave model each substep: the level set is taken directly from the canonical height field,
\(
\phi(\bm{x}, y) \;=\; y - \eta(\bm{x}, t),
\label{eq:fab_phi}
\)
with FLIP particles reseeded inside the band to maintain narrow-band density, and the cell-centered velocity is read off the layered bulk reconstruction derived in the remainder of this section.

\subsection{Bulk velocity potential from the canonical surface pair}
\label{sec:bulk_potential}

Recall that $\bm u = \nabla\Phi$ and $\Phi$ is harmonic in the fluid. Inside the strip $-d\le y\le \eta(\bm{x},t)$, $\Phi$ is uniquely determined by its surface trace $\psi(\bm{x},t) = \Phi|_{y=\eta}$ and the bottom no-flux condition $\partial_y\Phi|_{y=-d}=0$. Reconstructing $\Phi$ from $\psi$ on the moving surface is exactly the elliptic inverse to the Dirichlet--Neumann map of Section~\ref{sec:zakharov}, and the same Craig--Sulem expansion that produces $G(\eta)\psi$ produces, with one fewer FFT, the boundary value of $\Phi$ on the \emph{flat} reference plane $y=0$.
Specifically, writing $\Phi|_{y=0} = \sum_{j\ge 0} \Phi^{(j)}|_{y=0}$ as a Taylor series in homogeneous powers of $\eta$, the recursion of \citet{craig1993numerical} gives
\begin{align}
\Phi^{(0)}|_{y=0} &= \psi,
\label{eq:phi0}\\
\Phi^{(1)}|_{y=0} &= -\eta\,G_0\psi,
\label{eq:phi1}\\
\Phi^{(2)}|_{y=0} &= -\eta\,G_0\Phi^{(1)}|_{y=0} - \tfrac{1}{2}\eta^2\,B_0\psi.
\label{eq:phi2}
\end{align}
We retain the expansion through first order, $\Phi|_{y=0} \approx \psi - \varepsilon\,\eta\,G_0\psi$, computed via $3/2$-rule de-aliasing\rev{~\cite{orszag1971elimination}} exactly as in the time step. The second-order term \eqref{eq:phi2} contains $B_0$, whose $|\bm{k}|^2$ symbol amplifies any high-wavenumber forcing that the relaxation source (Section~\ref{sec:3d_to_2d}) injects into the wave state when the 3D box transmits a hull depression or a localized splash; we therefore truncate at first order to keep the boundary condition robust under arbitrary 3D forcing.

\subsection{Finite-depth harmonic kernels}
\label{sec:depth_kernels}

Once $\Phi|_{y=0}$ is in hand, the bulk potential at every depth follows from the harmonic extension into a flat strip with rigid bottom. The unique solution to $\Delta\Phi = 0$ on $\bm{x}\in\mathbb R^2,\ -d\le y\le 0$, with prescribed $\Phi|_{y=0}$ and $\partial_y\Phi|_{y=-d}=0$, separates per Fourier mode as
\begin{equation}
\hat\Phi(\bm{k},y) \;=\; \frac{\cosh\bigl(|\bm{k}|(y+d)\bigr)}{\cosh(|\bm{k}|d)}\;\hat\Phi(\bm{k},0).
\label{eq:harm_ext}
\end{equation}
For a query depth $y_i \in [0,d]$ measured downward from the flat reference (so the actual coordinate is $y = -y_i$), define
\begin{align}
C_i(\bm{k}) &\;=\; \frac{\cosh\bigl(|\bm{k}|(d - y_i)\bigr)}{\cosh(|\bm{k}|d)},
\label{eq:Ci}\\[2pt]
S_i(\bm{k}) &\;=\; |\bm{k}|\,\frac{\sinh\bigl(|\bm{k}|(d - y_i)\bigr)}{\cosh(|\bm{k}|d)}.
\label{eq:Si}
\end{align}
The two kernels carry the horizontal continuation factor and the vertical-derivative factor at depth $y_i$, respectively. Their limits encode the physical boundary conditions: at the surface ($y_i = 0$) we recover $C_0 = 1$ and $S_0 = G_0$, so the leading-order vertical surface velocity is exactly $G_0\psi$; at the bottom ($y_i = d$) we recover $C_d = 1/\cosh(|\bm{k}|d)$ and $S_d = 0$, enforcing the no-flux condition. In the deep-water limit $|\bm{k}|d\to\infty$, both kernels collapse to $e^{-|\bm{k}|y_i}$ (with $|\bm{k}|$ multiplied for $S_i$), recovering the exponential attenuation used by linear deep-water coupling schemes; our formulation \revv{strictly} generalizes \rev{this result} to finite depth.

For numerical robustness on float32 GPU arithmetic, we rewrite \eqref{eq:Ci}--\eqref{eq:Si} in their stable exponential form,
\begin{equation}
C_i = \frac{e^{-|\bm{k}|y_i} + e^{-|\bm{k}|(2d-y_i)}}{1 + e^{-2|\bm{k}|d}}, \quad
S_i = |\bm{k}|\,\frac{e^{-|\bm{k}|y_i} - e^{-|\bm{k}|(2d-y_i)}}{1 + e^{-2|\bm{k}|d}},
\label{eq:CiSi_stable}
\end{equation}
which avoids overflow when $|\bm{k}|d$ is large at high wavenumbers in shallow water. At $\bm{k}=\bm 0$ we set $C_i=1$, $S_i=0$ to suppress the spurious uniform-mode flow.

\begin{figure*}[!htbp]
    \centering
    \includegraphics[width=1.0\textwidth]{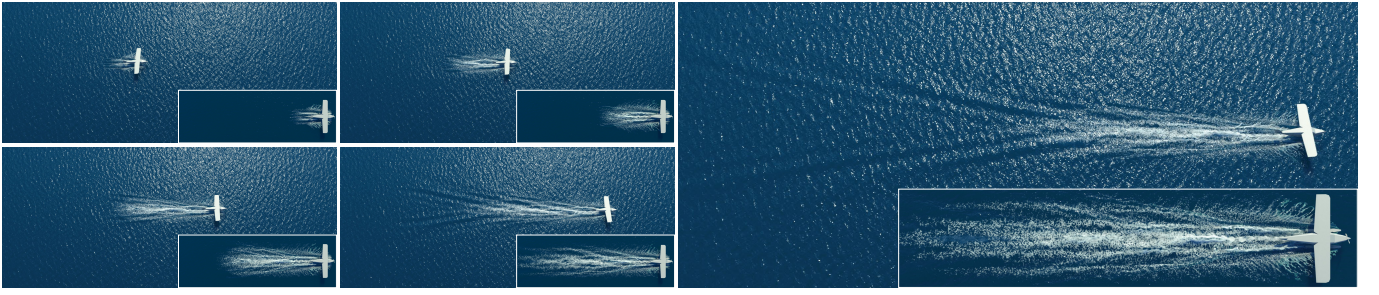}
    \caption{Seaplane landing on the ocean. A seaplane with a \SI{10}{\meter} wingspan, pitched \SI{9}{\degree} nose-up, approaches at \SI{8}{m/s}, touches down, and skims the surface at \SI{10}{m/s}, leaving a long trailing wake that transmits \revv{seamlessly}\rev{smoothly} across the 2D--3D interface. In each panel, the main image is the far view and the inset is the near view of the same moment. \rev{The wave model uses $\varepsilon=1$.}}
    \label{fig:seaplane}
\end{figure*}

\subsection{Velocity reconstruction at each depth layer}
\label{sec:depth_layers}

We sample the embedding box at $L_d$ depth layers, $y_i = i\,\Delta y_d$ for $i=0,\dots,L_d-1$, with the layer spacing $\Delta y_d$ chosen to match the box height. At each layer we obtain horizontal and vertical velocities from the spectral $\Phi|_{y=0}$ by
\begin{align}
\hat u_i(\bm{k}) &= ik_x\,C_i(\bm{k})\,\hat\Phi|_{y=0}(\bm{k}),
\label{eq:ui}\\
\hat w_i(\bm{k}) &= ik_z\,C_i(\bm{k})\,\hat\Phi|_{y=0}(\bm{k}),
\label{eq:wi}\\
\hat v_i(\bm{k}) &= S_i(\bm{k})\,\hat\Phi|_{y=0}(\bm{k}),
\label{eq:vi}
\end{align}
followed by inverse FFT and, when reflecting boundaries are active, cropping back to the physical $(M,N)$ grid. The cost is $L_d$ FFT pairs per substep, all on the same surface grid. We then trilinearly interpolate the layered velocity field onto the 3D box's lateral and bottom face cells, i.e., the FAB ring of cells that the NB-FLIP solver  consumes as a velocity boundary condition during pressure projection.
The layered reconstruction is summarized in Algorithm~\ref{alg:compute_volvel_from_2D}.

\begin{algorithm}[t]
\caption{ComputeVolumeVelFrom2D (bulk velocity from $(\eta,\psi)$ via the Craig--Sulem extension and finite-depth kernels)}
\label{alg:compute_volvel_from_2D}
\begin{flushleft}
    \textbf{Input}: surface state $(\eta,\psi)$ on wave grid, kernels $\{C_i,S_i\}_{i=0}^{L_d-1}$, padding mode \texttt{Pad}.\\
    \textbf{Output}: layered velocities $\{(u_i,v_i,w_i)\}_{i=0}^{L_d-1}$.
\end{flushleft}
\begin{algorithmic}[1]
\If{\texttt{Pad}}
    \State $(\eta,\psi) \gets$ \textbf{EvenExtend}$(\eta,\psi)$ \Comment{Sec.~\ref{sec:bc}}
\EndIf
\State $\hat\eta, \hat\psi \gets \textbf{FFT}(\eta), \textbf{FFT}(\psi)$
\State $\hat\Phi|_{y=0} \gets \hat\psi - \varepsilon\,\bigl(\eta\cdot G_0\psi\bigr)^{\wedge}_{3/2}$ \Comment{Eq.~\eqref{eq:phi1}}
\State $\hat\Phi|_{y=0} \gets \hat\Phi|_{y=0} \cdot \mathrm{taper}_{\mathrm{couple}}$ \Comment{Sec.~\ref{sec:bc_filter}}
\For{$i = 0,\dots,L_d-1$}
    \State $\hat u_i, \hat w_i \gets ik_x\,C_i\,\hat\Phi|_{y=0},\ ik_z\,C_i\,\hat\Phi|_{y=0}$
    \State $\hat v_i \gets S_i\,\hat\Phi|_{y=0}$
    \State $(u_i,v_i,w_i) \gets \textbf{IFFT}(\hat u_i,\hat v_i,\hat w_i)$
    \If{\texttt{Pad}} $(u_i,v_i,w_i) \gets$ \textbf{Crop}$(u_i,v_i,w_i)$ \EndIf
\EndFor
\end{algorithmic}
\end{algorithm}

%% file: 3Dto2D.tex
\section{3D-to-2D Coupling: Implicit Relaxation in Canonical Variables}
\label{sec:3d_to_2d}

The reverse half of the coupling closes the loop: at every substep, the 3D NS solution inside the embedding box must be transferred back to the wave model so that hull depressions, splashes, and other forced features born in the 3D regime appear \revv{seamlessly}\rev{smoothly} in the open-ocean surface field. The right panel of \autoref{fig:ocean_layers} sketches this direction: the 3D solver advances $(\bm u,\phi)$ inside the box and exposes its surface elevation $\eta_{3\mathrm D}$, which is then folded back into the wave-side $\eta$ through a relaxation zone supported on the box footprint. Two physical incompatibilities make this nontrivial. First, the wave model is canonical in $(\eta,\psi)$ but the 3D NS solver tracks a level set $\phi$ and a velocity field $\bm u$; there is no canonical $\psi$ directly available on the 3D side, so the wave model must reconstitute it. Second, naive overwriting of the wave state inside the box would create a discontinuity at the box edge that radiates spurious reflections back into the open ocean. We resolve the first issue by transferring only the surface elevation $\eta$ from the 3D solver and reconstructing the canonical $\psi$ on the wave side through \rev{an inverse DNO solve} $\psi=G(\eta)^{-1}v$, which converts the pair $(\eta,v)$ back into the canonical conjugate, so that no fluid-derived source acts on the dynamic equation for $\psi$. The second issue is handled by formulating the data transfer as an implicit relaxation source on the Zakharov right-hand side that decays exponentially through a smooth transition band.

\begin{figure*}[!htbp]
    \centering
    \includegraphics[width=1.0\textwidth]{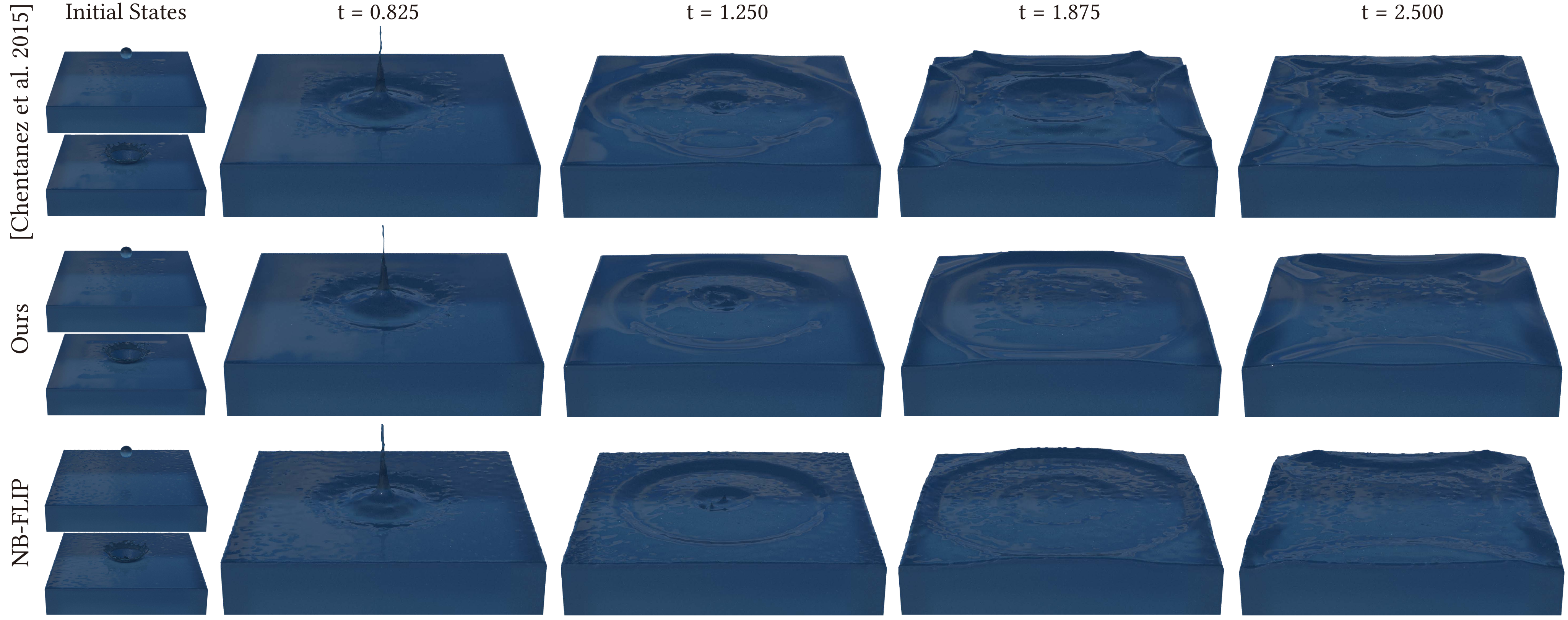}
    \caption{Crown Splash\revvv{ (Recolored to appear brighter)}. The crown splash scenario compares \citet{chentanez2015coupling} (top row), ours (middle row), and the 3D NB-FLIP reference (bottom row). Our method produces results that closely match the NB-FLIP reference, faithfully preserving crown splash dynamics and \rev{enabling smooth} wave transmission across the 2D--3D interface. In contrast, the method of \citet{chentanez2015coupling} exhibits clearly incorrect wave propagation in the 2D region, as the shallow water equations cannot reproduce the dispersive behavior of the deeper-water reference. \rev{Our wave model uses $\varepsilon=0.8$.}}
    \label{fig:ball_drop}
\end{figure*}

\subsection{Kinematic targets from the 3D state}
\label{sec:targets}

Let $\phi^{3\mathrm D}(\bm x, y, t)$ be the 3D level set at the current substep, with $y$ the vertical axis and $(x,z)$ the horizontal plane of the wave grid. We construct the elevation target $\eta^{\mathrm{target}}(x,z)$ on the wave grid by a column-wise scan of $\phi^{3\mathrm D}$, following the strategy of \citet{chentanez2015coupling}. For each wave-grid column $(i,k)$, we sweep the 3D cells from the bottom upward and count consecutive air/solid cells; once the running count reaches a small threshold $n_{\mathrm{air}}$ (we use $n_{\mathrm{air}}=3$), we identify the cell $n_{\mathrm{air}}$ steps below as the surface cell $j^{\star}$, and set
\begin{equation}
\eta^{\mathrm{target}}_{i,k} \;=\; \bigl(j^{\star}+\tfrac{1}{2}+o_y\bigr)\,\Delta x \;-\; \phi^{3\mathrm D}_{i,j^{\star},k},
\label{eq:eta_target_scan}
\end{equation}
i.e.\rev{,} the $y$-coordinate of the surface cell center, corrected by the signed-distance value at that cell as a sub-cell offset ($o_y$ is the box origin offset in $y$). To keep this target compatible with the wave grid, we then apply a few iterations of FAB-band padding plus a small Laplace smoothing to $\eta^{\mathrm{target}}$; this damps the high-frequency staircase noise that column-scanned level sets exhibit.



\subsection{Relaxation source on the Zakharov system}
\label{sec:relax_source}

With the elevation target $\eta^{\mathrm{target}}$ in hand, we drive the wave state toward the 3D state by augmenting the Zakharov system with a relaxation source on the canonical equations,
\begin{align}
\partial_t\eta &= G(\eta)\psi \;-\; \sigma(\bm x)\bigl(\eta - \eta^{\mathrm{target}}\bigr),
\label{eq:zk_relax_eta}\\
\partial_t\psi &= -g\eta + \mathcal N_\psi,
\label{eq:zk_relax_psi}
\end{align}
in which $\sigma(\bm x)\ge 0$ is a spatially graded relaxation rate (units $1/\mathrm{s}$) supported on the embedding-box footprint. We choose
\begin{equation}
\sigma(\bm x) \;=\; \sigma_{\max}\,\rho\!\left(\frac{\mathrm{dist}(\bm x,\partial\Omega_{3\mathrm D})}{w_{\mathrm b}}\right)\,\mathbf{1}_{\bm x\in\Omega_{3\mathrm D}},
\label{eq:sigma_def}
\end{equation}
where $\partial\Omega_{3\mathrm D}$ is the box edge, $w_{\mathrm b}$ is the transition-band width, $\rho:[0,1]\to[0,1]$ is a smooth ramp ($\rho(0)=0$, $\rho(1)=1$), and $\mathbf{1}_{\bm x\in\Omega_{3\mathrm D}}$ is the indicator of the box footprint, which restricts the support of $\sigma$ to the box interior. Outside the box, $\sigma(\bm x)\equiv 0$ and the wave evolves freely. In the transition band $\sigma$ rises from $0$ to $\sigma_{\max}$, and in the box interior $\sigma\equiv\sigma_{\max}$ pulls the wave state toward the 3D state at rate $\sigma_{\max}$.
Here, the rate $\sigma$ is a property of position, not of the wave field, so the source enters \eqref{eq:zk_relax_eta} \emph{linearly}; this is what makes it amenable to exact integration.

The source acts on the kinematic equation \eqref{eq:zk_relax_eta} alone because $(\eta,\psi)$ are canonically conjugate rather than two independent fields: the wave Hamiltonian already commits $\psi$'s evolution to $\eta$ through the dynamic equation \eqref{eq:zk_relax_psi}, so pinning $\eta$ to $\eta^{\mathrm{target}}$ already determines what $\psi$ must do, and adding an independent $\psi^{\mathrm{target}}$ source would \revv{over-determine} \rev{risk over-determining} the pair and \revv{break}\rev{breaking} the symplectic identity $v=G(\eta)\psi=\partial_t\eta$ inside the box. \rev{In practice such a fluid-derived target carries the NS solver's numerical viscosity and discretization artifacts, so it generally disagrees with the conjugate the wave Hamiltonian produces, and the mismatch radiates from the box edge.} We validate this choice empirically in \autoref{fig:relax_psi_ablation}: additionally driving $\psi$ toward a Helmholtz-decomposed target extracted from the 3D surface velocities produces visibly degraded results.

\subsection{Implicit integration of the relaxation source}
\label{sec:implicit_relax}

A naive explicit treatment $\eta_{n+1} = \eta_n - \sigma\Delta t\,(\eta_n-\eta^{\mathrm{target}})$ \rev{becomes} unstable for $\sigma_{\max}\Delta t > 2$, which would force a uselessly small $\Delta t$ when the box must absorb a violent feature. Because the source is linear in the deviation $\eta-\eta^{\mathrm{target}}$, the relaxation has a closed-form solution: holding $\eta^{\mathrm{target}}$ frozen across one substep, the relaxation flow is
\begin{equation}
\eta(t+\Delta t) \;=\; \eta^{\mathrm{target}} + \bigl(\eta(t) - \eta^{\mathrm{target}}\bigr)\,e^{-\sigma(\bm x)\Delta t},
\label{eq:eta_relax_exact}
\end{equation}
which is unconditionally stable, monotonically convergent, and exact at every $\sigma$. We compose this analytic decay with the wave step of Section~\ref{sec:time_step} via \rev{operator} splitting: first advance $(\eta,\psi)$ by $\Delta t$ under the wave Hamiltonian, then apply \eqref{eq:eta_relax_exact} pointwise.

\begin{figure*}[!htbp]
    \centering
    \includegraphics[width=1.0\textwidth]{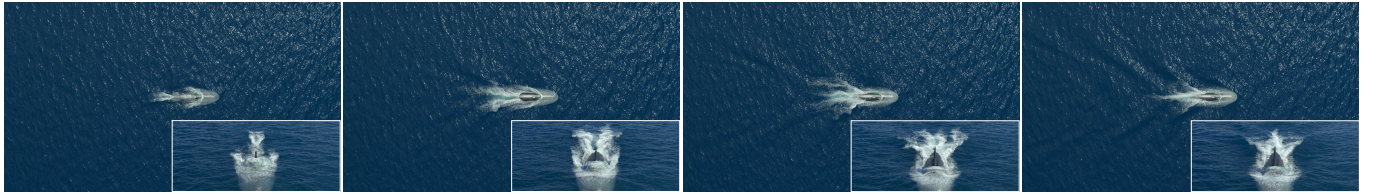}
    \caption{A surfacing submarine advancing forward and slowly rising to the ocean surface. In each panel, the main image is the far view and the inset is the near view of the same moment. \rev{The wave model uses $\varepsilon=0.3$.}}
    \label{fig:submarine}
\end{figure*}

\begin{figure*}[!htbp]
    \centering
    \includegraphics[width=1.0\textwidth]{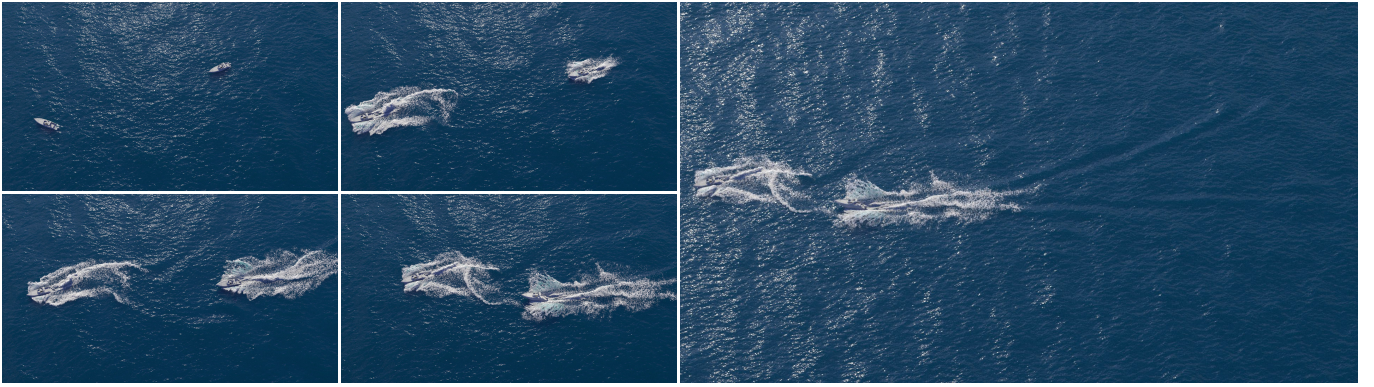}
    \caption{Two chasing boats on the ocean surface, producing long trailing wakes as they move through the water. The four images on the left show the near view, while the large image on the right shows the far view. \rev{The wave model uses $\varepsilon=1$.}}
    \label{fig:two_boat}
\end{figure*}

Equation \eqref{eq:eta_relax_exact} updates $\eta$ but says nothing about the wave's surface vertical velocity $v=G(\eta)\psi$, and we deliberately do not introduce a target $v^{\mathrm{target}}$ to drive it with. At every substep we preserve the wave's own $v$ across the $\eta$ update (the value $v=G(\eta_{\mathrm{step}})\psi_{\mathrm{step}}$ carried out of the Hamiltonian substep) and recompute the canonical conjugate from the post-relaxation pair $(\eta^{\mathrm{relaxed}}, v)$ via \rev{an inverse DNO solve},
\begin{equation}
\psi^{\,\mathrm{relaxed}}
=
\Bigl[G\bigl(\eta^{\,\mathrm{relaxed}}\bigr)\Bigr]^{-1} v,
\label{eq:psi_resync}
\end{equation}
see \eqref{eq:psi_inverse}. The triple $(\eta,\psi,v)$ is therefore \revv{canonically} self-consistent at the end of each substep.

We do not introduce a $v^{\mathrm{target}}$ because the information $v$ carries is already encoded in the time history of $\eta^{\mathrm{target}}$: the canonical relation $v=G(\eta)\psi$ together with the unmodified Hamilton evolution of $\psi$ means that pinning $\eta$ to $\eta^{\mathrm{target}}$ across substeps automatically pulls the wave's $v$ toward $\partial_t\eta^{\mathrm{target}}$, so a separate $v$ target would supply \rev{little} independent information. The single-boat experiment in \autoref{fig:single_boat} also \rev{supports} this choice: variant (c), which additionally relaxes $v$ toward a finite-difference target built from the 3D level set, introduces spurious reflections behind the stern, while variant (e), which relaxes $\eta$ alone, lets the outgoing wake pass through cleanly.



%% file: implementation.tex
\section{Implementation Details}
\label{sec:implementation}

\begin{figure*}[!htbp]
    \centering
    \includegraphics[width=1.0\textwidth]{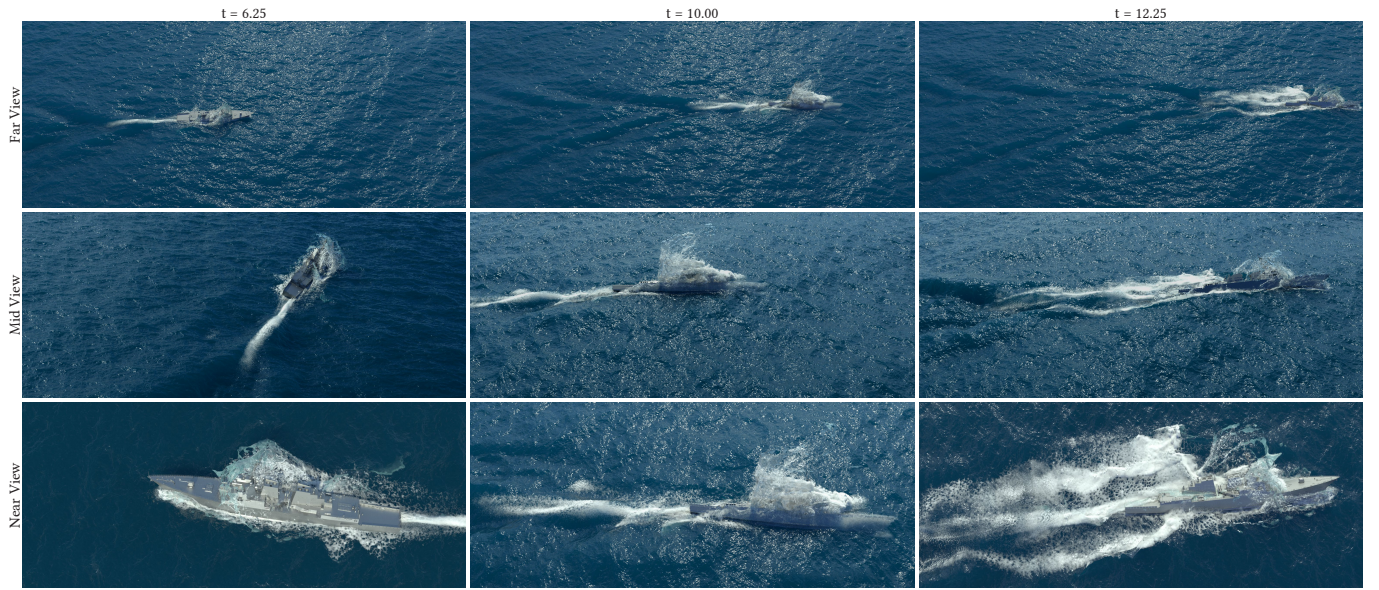}
    \caption{Battleship sailing through heavy waves under weak two-way solid--fluid coupling. The 2D wave domain measures \SI{120}{\meter}$\times$\SI{60}{\meter} and embeds a \SI{24}{\meter}$\times$\SI{12}{\meter}$\times$\SI{12}{\meter} 3D fluid box around the hull. Columns correspond to three time instants ($t = 6.25,\,10.00,\,12.25\,\text{s}$); the rows show, from top to bottom, the far, mid, and near views of the same scene. The simulation remains stable under the near-breaking background spectrum, and the 2D and 3D regions stay consistent \rev{ in the frames shown; residual boundary artifacts visible in the accompanying animation are discussed in \autoref{sec:validation}}. \rev{The wave model uses $\varepsilon=0.2$.}}
    \label{fig:battleship}
\end{figure*}

\begin{figure}[!htbp]
    \centering
    \includegraphics[width=0.5\textwidth]{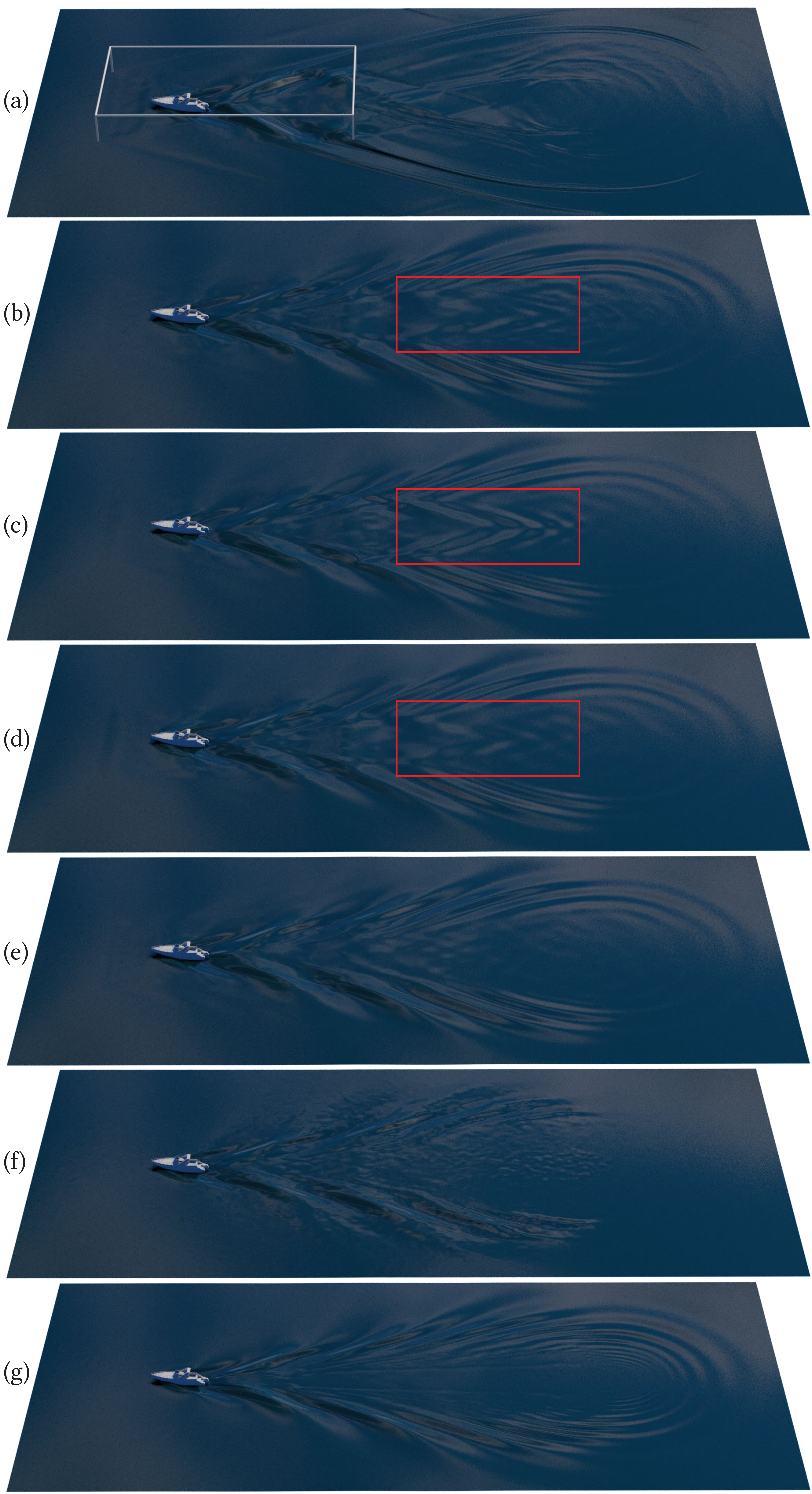}
    \caption{Single-boat wake comparison across seven methods.
    (a) \citet{chentanez2015coupling} \rev{(the white box marks the common 3D simulation region)}: the wake is qualitatively wrong, as the shallow water equations cannot resolve the dispersive Kelvin pattern.
    (b) Airy DK \cite{canabal2016dispersion} coupled via the FAB scheme \cite{stomakhin2017fluxed},
    (c) coupling through height and velocity relaxation, and
    (d) our wave model combined with the coupling scheme of \citet{chentanez2015coupling}
    all reproduce the feather-like wake \rev{\cite{rabaud2013ship}} but exhibit clear spurious reflections behind the stern (highlighted by the red boxes).
    (e) Our full method reproduces the feather-like Kelvin wake with minimal spurious reflection at the 2D--3D interface.
    (f) 3D NB-FLIP reference: the trailing wave at the far right is noticeably damped due to numerical dissipation of the 3D solver.
    (g) Pure 3D Eulerian level-set simulation.
    \rev{The variants built on our wave model, (c), (d), and (e), use $\varepsilon=1$.}}
    \label{fig:single_boat}
\end{figure}

Our method is implemented in Warp \cite{warp2022} with GPU acceleration. In 2D simulation, we utilize PyTorch's FFT module and padding functions to handle spectral operations and boundary conditions efficiently. Our 2D solver is compatible with standard ocean spectrum initializations. In some of our experiments, the initial height and velocity are \rev{generated from a TMA spectrum} with Donelan--Banner directional spreading, following \citet{horvath2015empirical}.
For 3D simulation, we implement NB-FLIP \cite{ferstl2016narrow} with weak two-way solid--fluid coupling following \cite{bridson2015fluid}. Solid objects are represented in voxelized cells. The simulation runs on staggered grids. Pressure projection is solved with a CUDA-based unsmoothed aggregated Algebraic Multigrid Preconditioned Conjugate Gradient solver \cite{sun2025lfm, shao2022fast}. Whitewater (spray, foam, and bubbles) shown in our results is simulated as a post-process using Houdini's built-in whitewater solver, driven by the velocity and surface fields produced by our simulation. \rev{The fine-scale ripples and shimmering visible on the ocean surface in our rendered results are small ocean-spectrum displacements added in Houdini at render time for visual richness; they are disabled in all quantitative comparisons.}

\paragraph{\rev{Masking residual coupling artifacts}}
\rev{While our coupling suppresses most interface artifacts, it does not eliminate them entirely. In our most demanding scene, the battleship in heavy seas (\autoref{fig:battleship}), the accompanying animation shows wavefronts reflecting from the rectangular boundary of the 3D box and propagating into the 2D domain. In production settings, such residual artifacts can be effectively obscured using two standard render-time layers. First, the ocean-spectrum displacements described above can be amplified to overlay additional procedural spectral detail on the simulated surface. Second, foam and whitewater can be emitted to further obscure the remaining artifacts.}

\subsection{De-aliased nonlinear evaluation}
\label{sec:dealias}

Because every $G_j$ for $j\ge 1$ is built from quadratic and higher-order pointwise products in $\eta$, $\nabla\eta$, $\nabla\psi$, $G_0\psi$, and $B_0\psi$, naive spectral evaluation aliases high-wavenumber content back into the resolved band and corrupts the nonlinear corrections within a few steps. We follow Orszag's $3/2$-rule\rev{~\cite{orszag1971elimination}}: every quadratic product in \eqref{eq:G1}, \eqref{eq:G2}, and \eqref{eq:dyn_truncated} is computed by zero-padding both factors from $(M,N)$ to $(\tfrac{3}{2}M,\tfrac{3}{2}N)$, multiplying in physical space on the padded grid, and truncating the result back to $(M,N)$. The padded grid is large enough that the highest aliased mode $|\bm k|+|\bm k'|=2|\bm k|_{\max}$ falls outside the retained band of the truncated product, so the surviving modes are exact. After de-aliasing we apply a smooth super-Gaussian taper $\exp[-(|\bm k|/k_{\mathrm c})^8]$ with $k_{\mathrm c} = \tfrac{2}{3}k_{\mathrm{Nyq}}$ to all nonlinear products. The taper is unity through most of the resolved band and prevents the high-wavenumber tail of $G_2$, whose symbol grows like $|\bm k|^2$, from amplifying numerical noise injected by the coupling boundary.

\subsection{Spectral bandlimiting of the boundary condition}
\label{sec:bc_filter}

The kernels in \eqref{eq:ui}--\eqref{eq:vi} act as spectral derivatives, so they amplify any high-wavenumber content in $\hat\Phi|_{y=0}$ linearly in $|\bm k|$. That tail is also the least reliable part of the wave field, since the truncated Craig--Sulem expansion loses accuracy near the wave grid's Nyquist. Feeding it unaltered into the 3D solver injects near-grid-scale forcing that the pressure projection then has to fight.
We therefore multiply $\hat\Phi|_{y=0}$ by a separable super-Gaussian taper $\exp[-(|k_\alpha|/(\tfrac{1}{2}k^\alpha_{\mathrm{Nyq}}))^8]$ for $\alpha\in\{x,z\}$ before applying \eqref{eq:ui}--\eqref{eq:vi}. The taper is unity through the lower half of the spectrum and decays smoothly to suppress the upper half, \rev{thereby preserving the resolved nonlinear wave content} while keeping the 3D pressure solve free of injected high-wavenumber noise.

%% file: time_int.tex
\section{Time Integration}
\label{sec:time_int}

\begin{figure*}[!t]
    \centering
    \includegraphics[width=1.0\textwidth]{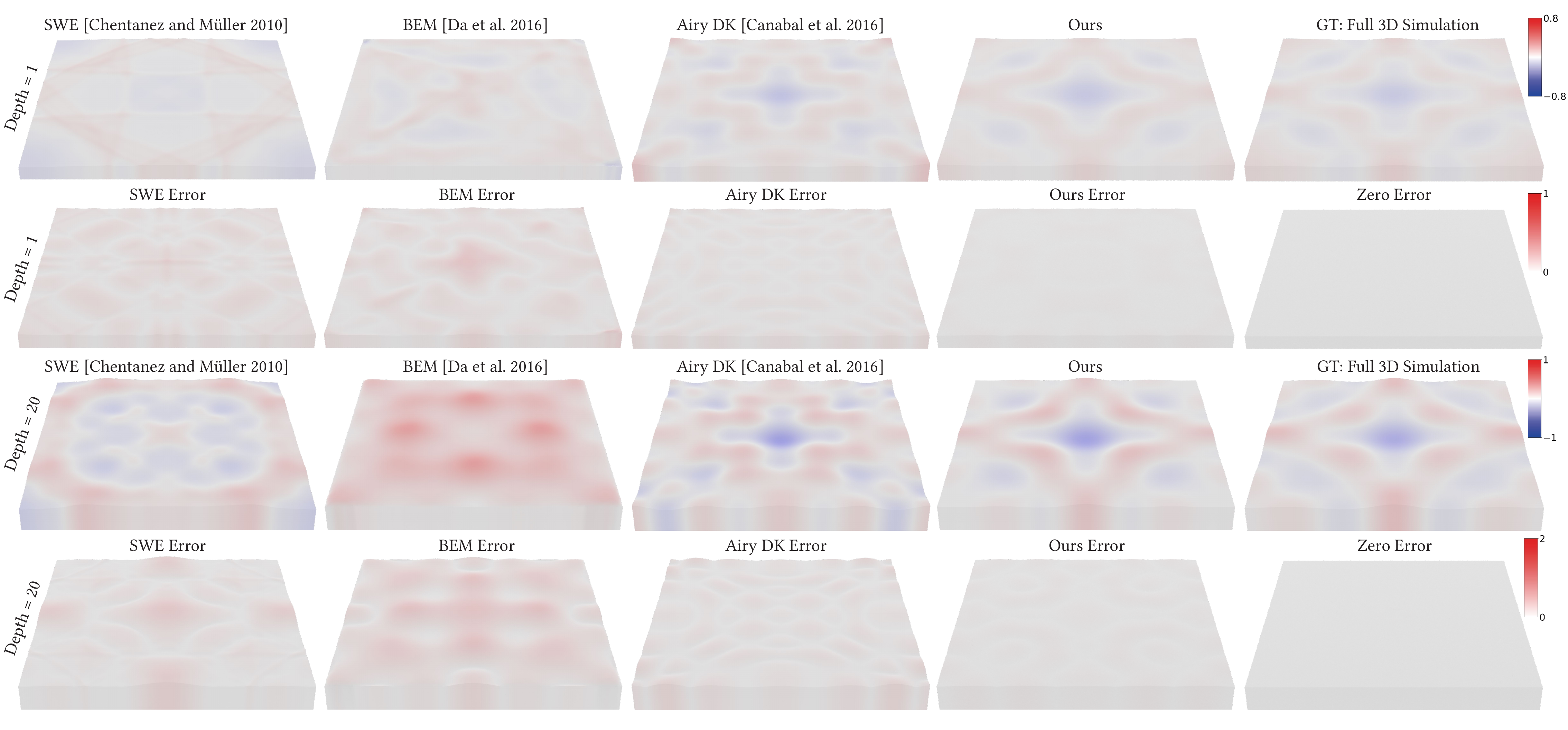}
    \caption{Wave propagation comparison in a 20-meter-wide tank under shallow (depth $=$ \SI{1}{\meter}) and deep (depth $=$ \SI{20}{\meter}) water conditions. For each depth, the upper row shows the rendered free surface produced by SWE~\cite{chentanez2010real}, BEM~\cite{da2016surface}, Airy DK~\cite{canabal2016dispersion}, our \rev{nonlinear} wave model \rev{ ($\varepsilon=1$)}, and a high-resolution 3D Eulerian level-set simulation that serves as the ground-truth reference; the lower row shows the corresponding per-cell absolute height error against this 3D reference. Our wave model agrees most closely with the 3D reference in both regimes.}
    \label{fig:wave_prop}
\end{figure*}

At each substep the 3D solver advances by $\Delta t_{3\mathrm D}$, set by an outer CFL on the 3D advection and solid motion. The 2D wave solver uses its own step $\Delta t_{2\mathrm D} \le \Delta t_{3\mathrm D}$: in most of our scenes we simply take $\Delta t_{2\mathrm D} = \Delta t_{3\mathrm D}$, but when finer resolution of the surface dynamics is desired we sub-cycle the wave solver with a smaller $\Delta t_{2\mathrm D}$ within one 3D step. The main algorithm is summarized below and outlined in \autoref{alg:main}. \rev{In the enumerated description that follows, the parenthetical labels refer to the corresponding line numbers of \autoref{alg:main}.}

\begin{enumerate}[leftmargin=*]
    \item \textbf{Compute Time Step (Step 1)}. Based on the CFL condition on the 3D velocity field and the solid motion, we compute the substep size $\Delta t$.

    \item \textbf{Advect 3D (Step 2)}. We perform the 3D level-set advection and velocity transport as in standard NB-FLIP, producing the tentative state $(\tilde\phi_{3\mathrm D}^{n+1}, \tilde{\bm u}_{3\mathrm D}^{n+1})$.

    \item \textbf{Update Solid (Step 3)}. We advance the rigid body by $\Delta t$ using its current linear and angular velocities and refresh the solid--fluid information carried by the 3D grid (solid SDF samples, solid velocity, and cell-type tags).

    \item \textbf{Prepare Targets (Steps 4--5)}. We build $\eta^{\mathrm{target}}$ from $\tilde\phi_{3\mathrm D}^{n+1}$ by the column-wise scan: sweep each column upward until $n_{\mathrm{air}}=3$ consecutive air/solid cells appear, then read the surface from $\tilde\phi_{3\mathrm D}^{n+1}$ at the cell $n_{\mathrm{air}}$ steps below (Eq.~\ref{eq:eta_target_scan}, Section~\ref{sec:targets}), followed by FAB-band padding and Laplace smoothing.


    \item \textbf{Advance 2D Simulation (Step 6)}. We advance the canonical wave state $(\eta,\psi)$ under the Zakharov Hamiltonian via the integrating-factor AB2 scheme \eqref{eq:if_ab2_eta}--\eqref{eq:if_ab2_psi} (Section~\ref{sec:wave}), producing the post-advance state $(\tilde\eta_{2\mathrm D}^{n+1}, \tilde\psi_{2\mathrm D}^{n+1})$.

    \item \textbf{Forward DNO Map (Step 7)}. We evaluate the post-advance surface vertical velocity $v = G\bigl(\tilde\eta_{2\mathrm D}^{n+1}\bigr)\,\tilde\psi_{2\mathrm D}^{n+1}$ via the forward DNO map \eqref{eq:v_forward} (Section~\ref{sec:state_recovery}).

    \item \textbf{Apply Relaxation (Step 8)}. We nudge $\tilde\eta_{2\mathrm D}^{n+1}$ toward $\eta^{\mathrm{target}}$ via the exact exponential decay \eqref{eq:eta_relax_exact} on $\eta$ alone, pointwise (Section~\ref{sec:implicit_relax}), producing the final height field $\eta_{2\mathrm D}^{n+1}$.

    \item \textbf{Reconstitute $\psi$ (Step 9)}. We recompute $\psi_{2\mathrm D}^{n+1}$ from $(\eta_{2\mathrm D}^{n+1}, v)$ through the inverse Dirichlet--Neumann map so that the canonical pair $(\eta_{2\mathrm D}^{n+1}, \psi_{2\mathrm D}^{n+1})$ is consistent with the relaxed height (Section~\ref{sec:implicit_relax}).

    \item \textbf{Embed FAB Height Field (Step 10)}. We set the level-set values of the 3D solver in the FAB ring from the post-relax 2D height field $\eta_{2\mathrm D}^{n+1}$ via $\phi = y - \eta$ (see Section~\ref{sec:fab_layer}); the interior of $\phi_{3\mathrm D}$ is left untouched.

    \item \textbf{Compute Volume Velocity from 2D (Step 11)}. We reconstruct the bulk velocity at each depth layer from the canonical 2D state using the Craig--Sulem lift to $y=0$ and the finite-depth harmonic kernels, as described in Sections~\ref{sec:bulk_potential}--\ref{sec:depth_layers} and summarized in \autoref{alg:compute_volvel_from_2D}.

    \item \textbf{Interpolate Velocity in FAB from Depth Layers (Step 12)}. We trilinearly interpolate the velocity within the FAB ring from the depth-layered velocities.

    \item \textbf{Advance 3D (Step 13)}. We follow standard NB-FLIP procedures to advance the 3D simulation, using the FAB-ring velocities as boundary conditions during pressure projection.
\end{enumerate}


\begin{algorithm}[t]
\caption{One substep of our coupled solver}
\label{alg:main}
\begin{algorithmic}[1]
\State $\Delta t \gets$ \textbf{ComputeTimeStep}() \Comment{outer 3D CFL}
\State $\tilde\phi_{3\mathrm D}^{n+1}, \tilde{\bm u}_{3\mathrm D}^{n+1} \gets$ \textbf{Advect3D}$\bigl(\Delta t, \bm u_{3\mathrm D}^{n}\bigr)$
\State $\mathcal{S}^{n+1} \gets$ \textbf{UpdateSolid}$\bigl(\Delta t, \mathcal{S}^{n}\bigr)$ \Comment{rigid-body pose}
\State $\tilde\eta_{\mathrm{samp3D}} \gets$ \textbf{Sample3DHeight}$\bigl(\tilde\phi_{3\mathrm D}^{n+1}\bigr)$ \Comment{Sec.~\ref{sec:targets}}
\State $\eta^{\mathrm{target}} \gets$ \textbf{Smooth}$\bigl(\tilde\eta_{\mathrm{samp3D}}\bigr)$ \Comment{Laplace smoothing}
\State $\tilde\eta_{2\mathrm D}^{n+1}, \tilde\psi_{2\mathrm D}^{n+1} \gets$ \textbf{Advance2D}$\bigl(\Delta t, \eta_{2\mathrm D}^{n}, \psi_{2\mathrm D}^{n}\bigr)$ \Comment{Sec.~\ref{sec:wave}}
\State $v \gets$ \textbf{ForwardDNO}$\bigl(\tilde\eta_{2\mathrm D}^{n+1}, \tilde\psi_{2\mathrm D}^{n+1}\bigr)$ \Comment{forward DNO, Eq.~\eqref{eq:v_forward}}
\State $\eta_{2\mathrm D}^{n+1} \gets$ \textbf{ApplyRelaxation}$\bigl(\Delta t, \tilde\eta_{2\mathrm D}^{n+1}, \eta^{\mathrm{target}}\bigr)$ \Comment{Sec.~\ref{sec:implicit_relax}}
\State $\psi_{2\mathrm D}^{n+1} \gets$ \textbf{ReconstitutePsi}$\bigl(\eta_{2\mathrm D}^{n+1}, v\bigr)$ \Comment{inverse DNO, Eq.~\eqref{eq:psi_inverse}}
\State $\phi_{3\mathrm D}^{n+1} \gets$ \textbf{EmbedFABHeight}$\bigl(\eta_{2\mathrm D}^{n+1}, \tilde\phi_{3\mathrm D}^{n+1}\bigr)$ \Comment{$\phi=y-\eta$ (FAB)}
\State $\bm u_{\mathrm{depth}}^{n+1} \gets$ \textbf{ComputeVolumeVelFrom2D}$\bigl(\eta_{2\mathrm D}^{n+1}, \psi_{2\mathrm D}^{n+1}\bigr)$ \Comment{\autoref{alg:compute_volvel_from_2D}}
\State $\bm u_{\mathrm{FAB}} \gets$ \textbf{InterpDepthLayerFAB}$\bigl(\bm u_{\mathrm{depth}}^{n+1}\bigr)$
\State $\bm u_{3\mathrm D}^{n+1} \gets$ \textbf{Advance3D}$\bigl(\Delta t, \bm u_{\mathrm{FAB}}, \tilde{\bm u}_{3\mathrm D}^{n+1}, \phi_{3\mathrm D}^{n+1}\bigr)$
\end{algorithmic}
\end{algorithm}

%% file: validation.tex
\section{Validation}
\label{sec:validation}

\paragraph{Experimental Setup} All experiments are conducted on a desktop with an NVIDIA GeForce RTX 4090. Implementation details of the 2D and 3D solvers are presented in \autoref{sec:implementation}. Relevant parameters and runtime performance are summarized in \autoref{statistics}. In the wave-propagation benchmark and the Stokes-wave accuracy study we \rev{report both HOS-2 and} HOS-3, while all other experiments use HOS-2, since this level is sufficient in the vast majority of our scenes and Order-2 is roughly $2\times$ faster than Order-3 per step (\autoref{tab:wave_prop_time_comp}).


\subsection{Comparisons and Validation Studies}

\paragraph{Wave Propagation}
\label{exp:wave}
To evaluate the accuracy of our wave model and our padding strategy, we simulated wave propagation in a square \SI{20}{\meter}$\times$\SI{20}{\meter} tank under shallow (\SI{1}{\meter}) and deep (\SI{20}{\meter}) water conditions with reflective boundary conditions on all four sides. In both cases the surface is initialized at rest ($v=0$) with a radially symmetric Gaussian height bump centered at the tank center, $\eta_0(x,z) = A\exp(-r^2/2\sigma^2)$ for $r<R$ and zero otherwise, where $r = \sqrt{(x-L/2)^2 + (z-L/2)^2}$, with width $\sigma=\SI{1}{\meter}$ and truncation radius $R=\SI{6}{\meter}$; the peak amplitude is $A=\SI{1}{\meter}$ in the shallow setting and $A=\SI{2}{\meter}$ in the deep setting. Each configuration is propagated for $2000$ frames at $\Delta t = \SI{0.01}{\second}$. As shown in Figure~\ref{fig:wave_prop}, we compared a total of four 2D wave models, i.e., the SWE~\cite{chentanez2010real} and the BEM~\cite{da2016surface} used in previous coupling methods~\cite{thurey2006animation, chentanez2015coupling, huang2021ships}, the Airy Dispersion Kernel (DK) model~\cite{canabal2016dispersion}, and our \revv{Hamiltonian}\rev{nonlinear} wave model, against a high-resolution 3D Eulerian level-set simulation that serves as the ground-truth reference. We found that our \revv{Hamiltonian}\rev{nonlinear} wave model produces results most consistent with the 3D reference, and correctly captures wave reflection.

We quantify these differences in \autoref{tab:wave_prop_err}. For a fair comparison, every method's output is resampled onto the same $128\times128$ height field before being compared with the reference; for BEM, which represents the surface as a mesh, we cast vertical rays downward from this $128\times128$ grid and take the first ray--mesh intersection as the per-cell height. For each frame we compute the absolute height error against the reference and reduce it to its spatial mean and max; we then average each of these quantities over the 2000 frames of the simulation, yielding the time-averaged mean and max errors reported in the table. Compared with SWE and BEM, our model reduces the time-averaged mean error by roughly $4\times$ in the shallow-water case and by $3$--$5\times$ in the deep-water case; even compared with the strongest baseline (Airy DK), it still cuts the mean error by \revv{about $2\times$ in both regimes}\rev{about $2.4\times$ in the shallow-water case and $1.7\times$ in the deep-water case}, with similar improvements on the max error. \rev{We also tested a strengthened Airy DK baseline that multiplies the dispersion relation by the amplitude-dependent factor $(1+a^2k^2)$. The correction improves the Airy DK baseline (row ``Airy DK + dispersion fix'' in \autoref{tab:wave_prop_err}), lowering its mean error by roughly $10\%$ in both regimes, but the corrected baseline still incurs about $1.5$--$2.1\times$ our mean error.} Runtime comparisons \rev{among SWE, BEM, Airy DK, and our Order 2 and 3 models} are listed in \autoref{tab:wave_prop_time_comp}; our wave model runs about 1000$\times$ faster than BEM (though it could be further optimized). \rev{While our model is more expensive than the linear-wave-based SWE and Airy DK solvers, it delivers the accuracy improvements reported in \autoref{tab:wave_prop_err} and remains substantially cheaper than BEM.} The 2D grid spacings used were \( \Delta x = \frac{L}{128} \), \( \frac{L}{128} \), \( \frac{L}{128} \), and \( \frac{L}{64} \) for the SWE, Airy DK, our wave model, and BEM, respectively, matching the grid spacing ratios used in \citet{huang2021ships}, while the 3D Eulerian reference uses a much finer grid spacing of \( \Delta x = \frac{L}{256} \).

\begin{figure}[!htbp]
    \centering
    \begin{subfigure}[b]{0.5\linewidth}
        \centering
        \includegraphics[width=\linewidth]{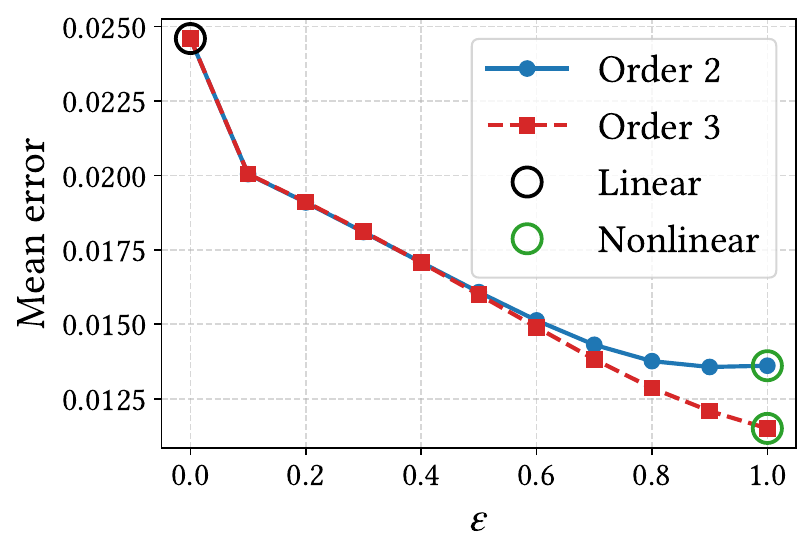}
        \caption{Mean error}
        \label{fig:wave_prop_mean_err}
    \end{subfigure}%
    \begin{subfigure}[b]{0.5\linewidth}
        \centering
        \includegraphics[width=\linewidth]{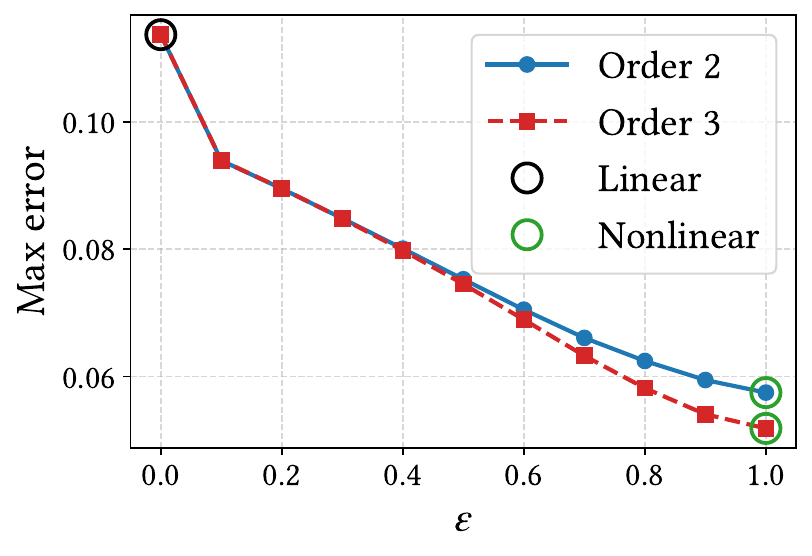}
        \caption{Max error}
        \label{fig:wave_prop_max_err}
    \end{subfigure}
    \caption{Wave height error in the shallow water wave propagation experiment, plotted against the 3D Eulerian reference as a function of the nonlinearity parameter $\varepsilon$, for both Order 2 and Order 3 HOS truncations. Hollow black circles mark the linear case ($\varepsilon=0$); hollow green circles mark the fully nonlinear case ($\varepsilon=1$). The error decreases monotonically as $\varepsilon$ grows, confirming that the higher-order HOS terms are correctly oriented; even a small $\varepsilon$ already removes a large fraction of the linear-model error. Order 2 and Order 3 are nearly indistinguishable for $\varepsilon\lesssim0.6$ and diverge beyond, with Order 3 attaining a noticeably lower error at $\varepsilon=1$.}
    \label{fig:wave_prop_err}
\end{figure}

To further probe the role of the nonlinear correction, we sweep the nonlinearity parameter $\varepsilon$ from $0$ (linear Airy) to $1$ (fully nonlinear HOS) in the shallow-water setting, using both the Order 2 and Order 3 HOS truncations, and measure the wave-height error against the same 3D reference. As shown in \autoref{fig:wave_prop_err}, the error decreases monotonically as $\varepsilon$ grows, confirming that the higher-order HOS terms are correctly oriented and pull the solver toward the 3D reference rather than away from it. The drop between $\varepsilon=0$ and $\varepsilon=0.1$ is already substantial, indicating that even a mild amount of nonlinearity removes a large fraction of the linear-model error. Order 2 and Order 3 are nearly indistinguishable for $\varepsilon \lesssim 0.6$, but the two curves separate beyond this point, with Order 3 attaining a noticeably lower error at $\varepsilon=1$.


\begin{table}
\centering\small
\caption{Runtime comparison of the wave propagation experiment.}
\begin{tabular}{l@{\hspace{2mm}}c@{\hspace{2mm}}c@{\hspace{2mm}}c}
\hlineB{3}
Method  &
\makecell{Deep Water Time (ms)} &
\makecell{Shallow Water Time (ms)} \\
\hlineB{2}
BEM & 8693.7 & 25973.8\\
\hlineB{2}
\rev{SWE} & \rev{1.355} & \rev{1.375}\\
\hlineB{2}
\rev{Airy DK} & \rev{1.081} & \rev{1.157}\\
\hlineB{2}
Ours (Order 2) & 6.352 & 6.167\\
\hlineB{2}
Ours (Order 3) & 13.52 & 13.43\\
\hlineB{3}
\end{tabular}
\label{tab:wave_prop_time_comp}
\end{table}

\begin{table}
\centering\small
\caption{Wave height error comparison of the wave propagation experiment. For each method, the first row is the mean error and the second row is the max error, each obtained by computing the spatial mean (resp.\ max) of the absolute height error on a $128\times128$ grid at every frame and then averaging over the 2000 frames of the simulation.}
\begin{tabular}{l@{\hspace{3mm}}c@{\hspace{3mm}}c@{\hspace{3mm}}c}
\hlineB{3}
Method & Error & Shallow & Deep \\
\hlineB{2}
\multirow{2}{*}{SWE \cite{chentanez2010real}} & Mean & 0.048202 & 0.096199 \\
                     & Max  & 0.210926 & 0.422074 \\
\hline
\multirow{2}{*}{BEM \cite{da2016surface}} & Mean & 0.051720 & 0.170493 \\
                     & Max  & 0.234508 & 0.466547 \\
\hline
\multirow{2}{*}{Airy DK \cite{canabal2016dispersion}} & Mean & 0.027666 & 0.054496 \\
                         & Max  & 0.126102 & 0.298801 \\
\hline
\multirow{2}{*}{\rev{Airy DK + dispersion fix}} & \rev{Mean} & \rev{0.024655} & \rev{0.049221} \\
                         & \rev{Max}  & \rev{0.113956} & \rev{0.280982} \\
\hline
\multirow{2}{*}{Ours (Order 2)} & Mean & 0.013607 & \textbf{0.032774} \\
                                & Max  & 0.057469 & 0.174430 \\
\hline
\multirow{2}{*}{Ours (Order 3)} & Mean & \textbf{0.011501} & 0.034247 \\
                                & Max  & \textbf{0.051843} & \textbf{0.161604} \\
\hlineB{3}
\end{tabular}
\label{tab:wave_prop_err}
\end{table}

\begin{figure}[!htbp]
    \centering
    \begin{subfigure}[b]{0.5\linewidth}
        \centering
        \includegraphics[width=\linewidth]{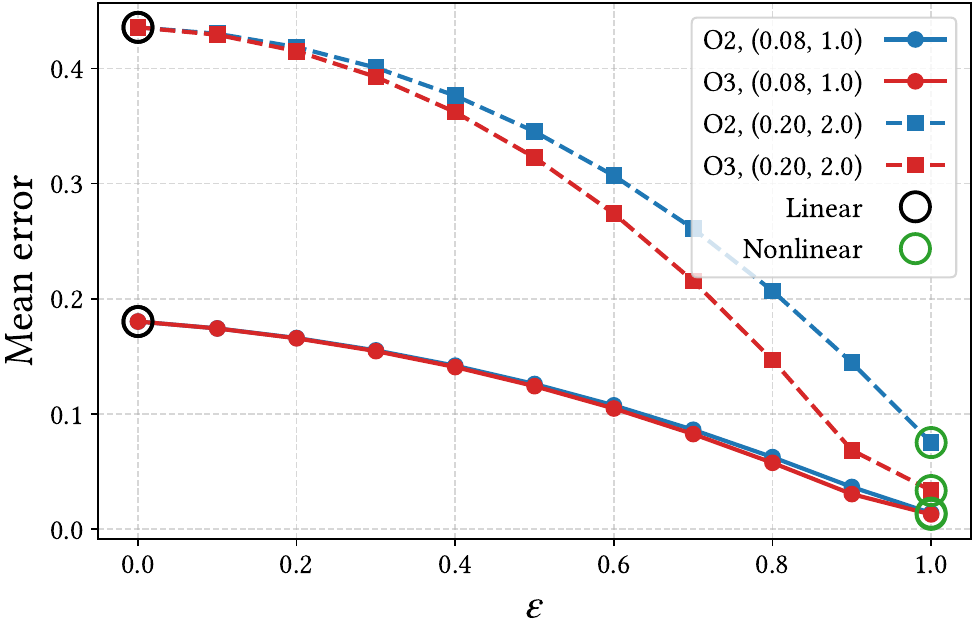}
        \caption{Mean error}
        \label{fig:eps_sweep_mean_err}
    \end{subfigure}%
    \begin{subfigure}[b]{0.5\linewidth}
        \centering
        \includegraphics[width=\linewidth]{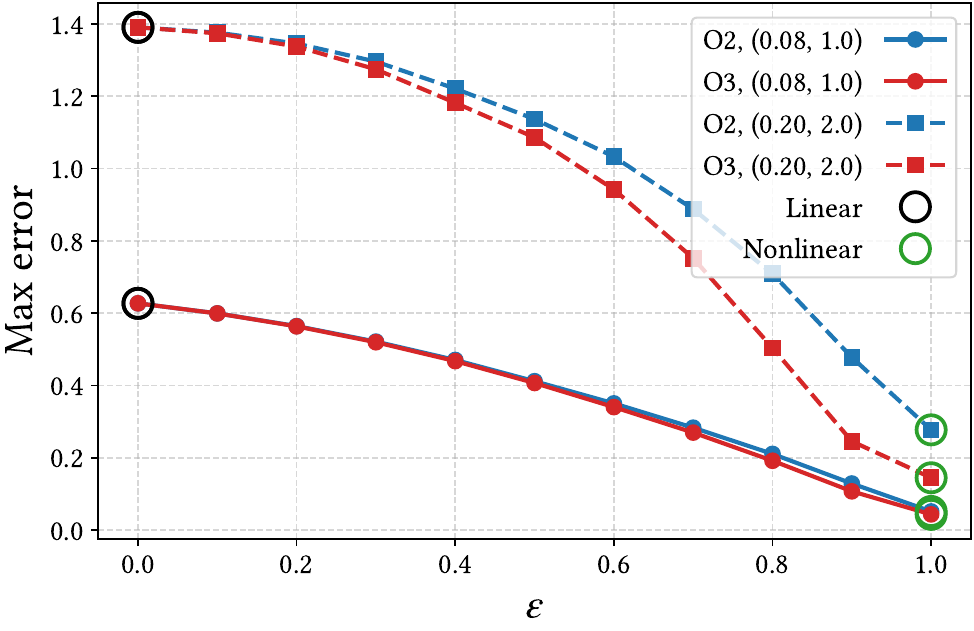}
        \caption{Max error}
        \label{fig:eps_sweep_max_err}
    \end{subfigure}
    \caption{Stokes wave propagation error against the analytical second-order Stokes solution as a function of the nonlinearity parameter $\varepsilon$. Four curves correspond to the two HOS truncation orders (O2, O3) and the two Stokes initial conditions parameterized by $(ka,\,kh)$. Hollow black circles mark the linear endpoint ($\varepsilon=0$); hollow green circles mark the fully nonlinear endpoint ($\varepsilon=1$).}
    \label{fig:eps_sweep_err}
\end{figure}

\paragraph{Stokes Wave Accuracy under Varying Nonlinearity}
We assess the accuracy of our HOS wave solver against the analytical second-order Stokes wave (with the third-order frequency correction) on a periodic 1D domain of length $L = 2\pi$ through two complementary tests.
\rev{Explicitly, the reference is the classical second-order Stokes profile \cite{dean1991water}}
\begin{equation*}
\eta_{\text{St}}(x,t) = a\cos\theta + \frac{k a^{2}\,\bigl(3-\tanh^{2}(kh)\bigr)}{4\tanh^{3}(kh)}\,\cos 2\theta
\end{equation*}
\rev{with $\theta = kx - \omega t$, whose frequency follows the third-order Stokes dispersion relation \cite{whitham2011linear} $\omega^{2} = gk\tanh(kh)\,(1 + \delta)$ with}
\begin{equation*}
\delta = \frac{(ka)^{2}\,\bigl(9 - 10\tanh^{2}(kh) + 9\tanh^{4}(kh)\bigr)}{8\tanh^{4}(kh)}.
\end{equation*}
\rev{The correction $\delta$ is the amplitude-dependent Stokes frequency shift, and dropping it recovers the linear relation $\omega^{2} = gk\tanh(kh)$.}

\textit{Long-time phase fidelity.} We initialize four mild Stokes regimes\rev{, i.e., weakly nonlinear, non-breaking Stokes waves,} with $(ka,\,kh) \in \{(0.08, 1.0),\,(0.10, 1.5),\,(0.12, 1.8),\,(0.13, 2.0)\}$ and propagate each for $18$, $14$, $12$, and $10$ linear periods, respectively. \autoref{fig:illustration} overlays the final-time surface elevation produced by the linear ($\varepsilon = 0$)\rev{, intermediate ($\varepsilon = 0.5$),} and fully nonlinear ($\varepsilon = 1$, HOS-3) solvers with the analytical Stokes profile. The linear solver (equivalent to \cite{canabal2016dispersion}) progressively drifts out of phase across all four regimes because it lacks the amplitude-dependent Stokes frequency shift \rev{ defined above}, while the \rev{fully nonlinear} HOS-3 solver remains phase-locked to the analytical curve. \rev{The intermediate $\varepsilon=0.5$ curve lies between the two, but closer to the linear one.}

\textit{Error sweep over $\varepsilon$.} To examine how the nonlinearity parameter $\varepsilon$ controls accuracy, we sweep $\varepsilon \in \{0, 0.1, \ldots, 1.0\}$ across two HOS truncation orders (Order 2 and Order 3) and two distinct initial conditions with $(ka,\,kh) \in \{(0.08, 1.0),\,(0.20, 2.0)\}$. We propagate each configuration for ten linear periods and compare against the analytical Stokes solution frame by frame. The spatio-temporal error field $|\eta_\text{sim} - \eta_\text{Stokes}|/a$ is reduced to its mean and max, as shown in \autoref{fig:eps_sweep_err}. The error decreases monotonically as $\varepsilon$ grows from $0$ to $1$, confirming that the higher-order DNO terms steer the linear solver toward the analytical Stokes wave. For the milder initial condition, Order 2 and Order 3 are nearly indistinguishable, indicating that the second-order kinematic boundary condition already captures the relevant nonlinear physics; for the steeper case, Order 3 noticeably outperforms Order 2, justifying the inclusion of the third-order term $G_2$ when handling steep waves.

\paragraph{Continuous Wavetrain Across the Coupling Interface}
To stress-test the 2D--3D coupling under sustained periodic forcing, we drive a steady second-order Stokes wavetrain through an embedded Eulerian fluid box, as shown in \autoref{fig:wavetrain}. A 1D wave domain of length \SI{3}{\meter} and still-water depth \SI{0.4}{\meter} hosts a \SI{2}{\meter}$\times$\SI{0.8}{\meter} fluid box centered at the origin. A wavemaker zone at the left edge of the wave domain softly relaxes $(\eta,\,v)$ toward the analytical Stokes solution with steepness $ka=0.3$ and wavelength \SI{0.8}{\meter}, generating a right-going train; the entire right side of the wave domain acts as a sponge that absorbs outgoing waves before they wrap around the periodic boundary. The fluid box is initialized with still water and exchanges $(\eta,\,\psi)$ with the 1D solver through the relaxation-zone protocol.

We use this setup to contrast the linear ($\varepsilon=0$, equivalent to \cite{canabal2016dispersion}) and fully nonlinear ($\varepsilon=1$) wave models when both are coupled to the same 2D Eulerian fluid box. \autoref{fig:wavetrain} shows three representative snapshots at $t\approx\SI{2.65}{\second},\,\SI{3.75}{\second},\,\SI{5.0}{\second}$ (about $3.7$, $5.2$, and $7.0$ wave periods). The red boxes mark the wave free-propagation region, where the 1D wave solver carries the train on its own. With the linear model this region develops visible kinks, the surface is no longer smooth, and the Stokes crest--trough asymmetry is entirely missing. The nonlinear model, by contrast, produces a smooth Stokes train whose shape inside the red boxes is continuous with the surface shape inside the 2D fluid box, with no visible discontinuity at the coupling interface. This indicates that the natively nonlinear 2D \rev{NS} solver is incompatible with the \rev{1D} linear Airy wave model \rev{at this steepness ($ka=0.3$)}: only the HOS wave model can match the 2D solver's surface dynamics across the coupling\rev{; for gentler, nearly linear wavetrains the two models would agree and the cheaper Airy model would suffice}. \rev{We note that this 1D setting has no transverse geometric spreading. In a full 2D wave domain, localized outgoing disturbances generally weaken as their wavefront expands, reducing nonlinear effects with distance.}


\begin{figure*}[!t]
    \centering
    \includegraphics[width=1.0\textwidth]{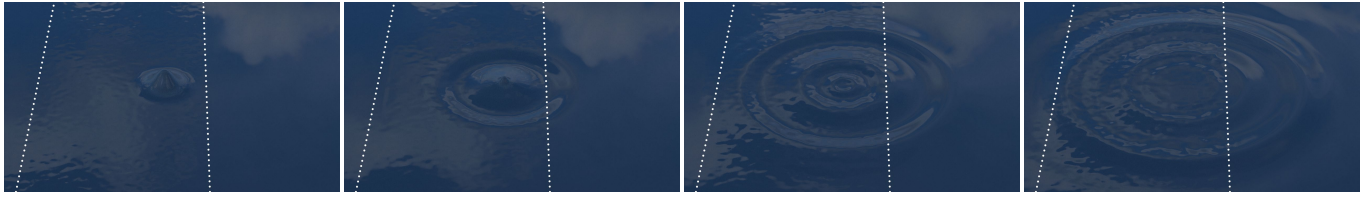}
    \caption{Dispersion matching test. The wave propagates from the 3D domain (bounded by dotted lines) into 2D regions. The wave retains a symmetric shape throughout its propagation, demonstrating consistent dispersion relations between the 3D and 2D simulations under our two-way coupled scheme. \rev{The wave model uses $\varepsilon=1$.}}
    \label{fig:dispersion_test}
\end{figure*}
\paragraph{Dispersion Matching Test}

The 2D and 3D \rev{simulations} may exhibit distinct numerical viscosity and dispersion characteristics. As noted by \cite{schreck2022coupling}, such differences can potentially lead to asymmetric wave propagation and spurious reflections at the coupling boundary. To investigate this, we conducted an experiment in a \SI{5}{\meter} water tank with a depth of \SI{1}{\meter}, where a wave packet with a peak amplitude of \SI{0.2}{\meter} was initialized inside the 3D domain. The results are shown in \autoref{fig:dispersion_test}, demonstrating strong agreement between our 2D and 3D models: the wave maintained a highly symmetric shape throughout its propagation, and no artificial reflections were observed at the 2D--3D interface.

To highlight the importance of dispersion matching, we compared our method against three prior coupling schemes in \autoref{fig:dispersion_compare}: \citet{chentanez2015coupling}, \citet{schreck2022coupling}, and FAB \cite{stomakhin2017fluxed} combined with the Airy DK wave model \cite{canabal2016dispersion}. In this combined baseline, the outer 2D wave field is advanced with the Airy DK model \cite{canabal2016dispersion}, the 2D-to-3D direction of the coupling follows the FAB scheme \cite{stomakhin2017fluxed}, and the 3D-to-2D direction is handled with the procedure of \citet{chentanez2015coupling}; since the 3D-to-2D step is essentially standard and shared across most coupling methods, we label this baseline simply as FAB + Airy DK. All methods are driven by the same 3D NB-FLIP solver under identical initial conditions. \rev{ The method of \citet{chentanez2015coupling} develops pronounced left--right asymmetry. For \citet{schreck2022coupling}, we recalibrate their dispersion-law rescaling factor for our 3D solver configuration and find that $d=0.8$ gives the best result. This largely removes the left--right asymmetry, although the 2D and 3D regions remain visibly less consistent than with our method. We note that \citet{schreck2022coupling} report no asymmetry under their original simulation settings. The FAB + Airy DK combination shows no visible asymmetry, but produces visibly deeper waves inside the 3D region than in the surrounding 2D field. Our method yields a wave field that is virtually identical inside and outside the 3D region, with a smoother 3D region.} \rev{In terms of runtime, the wave-solver cost in this comparison is \SI{13.3}{\milli\second} per substep for our method, \SI{27.5}{\milli\second} for \citet{schreck2022coupling}, \SI{3.7}{\milli\second} for FAB + Airy DK, and \SI{2.3}{\milli\second} for \citet{chentanez2015coupling}.}

\begin{table*}[htbp]
\centering
\caption{Specifications and statistics of the simulations. \rev{``Simulation Substeps'' refers to 3D substeps.}}
\small
\resizebox{\textwidth}{!}{
\begin{tabular}{c|c|c|c|c|c|c|c|c|c|c}
\hlineB{3}
Case & Res 2D & Res 3D & $\Delta x_\text{2D}$~[\SI{}{\meter}] & $\Delta x_\text{3D}$~[\SI{}{\meter}] & \makecell[c]{Frame \\ Count} & \makecell[c]{Frame \\ Time Step~[\SI{}{\second}]} & \makecell[c]{Simulation \\ Substeps} & $T_\text{substep}$~[\SI{}{\milli\second}] & CFL & $\varepsilon$ \\
\hlineB{2}
Dispersion Matching      & 256$\times$256   & 256$\times$64$\times$512    & 0.020  & 0.0078 & 200 & 1/80 & 1133  & 87.8  & 0.25 & 1.0 \\ 
\hline
Crown Splash             & 128$\times$128   & 128$\times$256$\times$128   & 0.0094 & 0.0063 & 300 & 1/80 & 665   & 58.4  & 1.0  & 0.8 \\ 
\hline
Single Boat (NB-FLIP)    & /                & 1152$\times$96$\times$576   & /      & 0.063  & 200 & 1/20 & 1273  & 363   & 1.0  &   /  \\ 
\hline
Single Boat              & 576$\times$288   & 384$\times$96$\times$192    & 0.13   & 0.063  & 200 & 1/20 & 1073  & 85.0  & 1.0  & 1.0 \\ 
\hline
Seaplane                 & 1024$\times$256  & 768$\times$128$\times$256   & 0.156  & 0.047  & 240 & 1/20 & 3951  & 197.2 & 1.0  & 1.0 \\ 
\hline
Battleship               & 512$\times$256   & 512$\times$256$\times$256   & 0.234  & 0.047  & 240 & 1/20 & 3408  & 160.8 & 1.0  & 0.2 \\ 
\hline
Submarine                & 512$\times$256   & 512$\times$128$\times$256   & 0.156  & 0.047  & 360 & 1/30 & 2584  & 99.7  & 1.0  & 0.3 \\ 
\hline
Two Chasing Boats        & 512$\times$256   & 512$\times$128$\times$256   & 0.35   & 0.047  & 300 & 1/30 & 3396  & 268.8 & 1.0  & 1.0 \\ 
\hline
Pond                     & 1024$\times$1024 & 512$\times$256$\times$512   & 0.0020 & 0.0018 & 300 & 1/30 & 14447 & 326.0 & 1.0  & 0.2 \\ 
\hlineB{2}
\end{tabular}
}
\label{statistics}
\end{table*}

\paragraph{Crown Splash}

A water ball with a radius of \SI{0.05}{\meter} falls from a height of \SI{0.2}{\meter} into a tank measuring \SI{1.2}{\meter} wide and \SI{0.6}{\meter} deep, as shown in \autoref{fig:ball_drop}. We compare our coupling scheme against a 3D NB-FLIP reference and \citet{chentanez2015coupling}. We apply reflected padding to the boundaries of the 2D solver to enforce solid boundaries. Our coupling scheme produces results nearly identical to the NB-FLIP ground truth, enabling smooth wave transmission across the 2D--3D interface and accurate wave propagation throughout the whole domain. The method of \citet{chentanez2015coupling}, on the other hand, shows clearly incorrect wave propagation in the 2D region: since it relies on the shallow water equations, it fails to capture the dispersive wave dynamics.

\begin{figure}[!htbp]
    \centering
    \begin{minipage}{0.25\textwidth}\centering\small Ours (w/o $\psi$ relaxation)\end{minipage}%
    \begin{minipage}{0.25\textwidth}\centering\small With $\psi$ relaxation\end{minipage}\\[2pt]
    \includegraphics[width=0.5\textwidth]{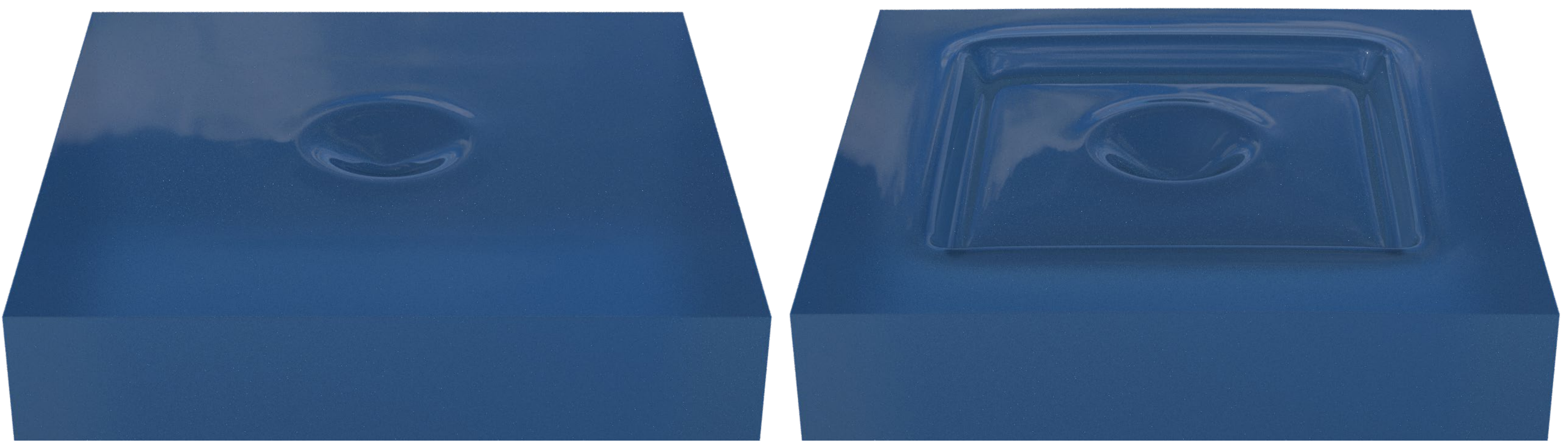}
    \caption{Ablation on relaxing $\psi$, on the Crown Splash setup. The right panel additionally relaxes $\psi$ toward a target obtained from a Helmholtz decomposition of the 3D surface velocity field, while the left panel is our method. Relaxing $\psi$ produces an obvious abnormal depression in the 3D region.}
    \label{fig:relax_psi_ablation}
\end{figure}

\paragraph{$\psi$ Relaxation Ablation}
\label{exp:relax_psi_ablation}
\rev{On the same Crown Splash setup, we ablate the asymmetric source placement of Section~\ref{sec:relax_source} by additionally relaxing $\psi$ toward a Helmholtz-decomposed target extracted from the 3D surface velocity field. As shown in \autoref{fig:relax_psi_ablation}, this symmetric variant produces spurious surface artifacts inside and around the embedding box.}

\paragraph{\rev{Filter and Time-Integrator Ablation}}
\rev{We further ablate two components of the wave solver. First, disabling the low-pass state filter \eqref{eq:state_filter} causes the Crown Splash simulation to quickly become unstable. Second, replacing the IF-AB2 integrator (Section~\ref{sec:time_step}) with a first-order integrating-factor (exponential) Euler step increases the mean/max wave height error in the shallow-water wave-propagation experiment from 0.011501/0.051843 to 0.012752/0.053980 (cf.\ \autoref{tab:wave_prop_err}).}


\paragraph{Single Boat}

\begin{figure}[!htbp]
    \centering
    \includegraphics[width=0.5\textwidth]{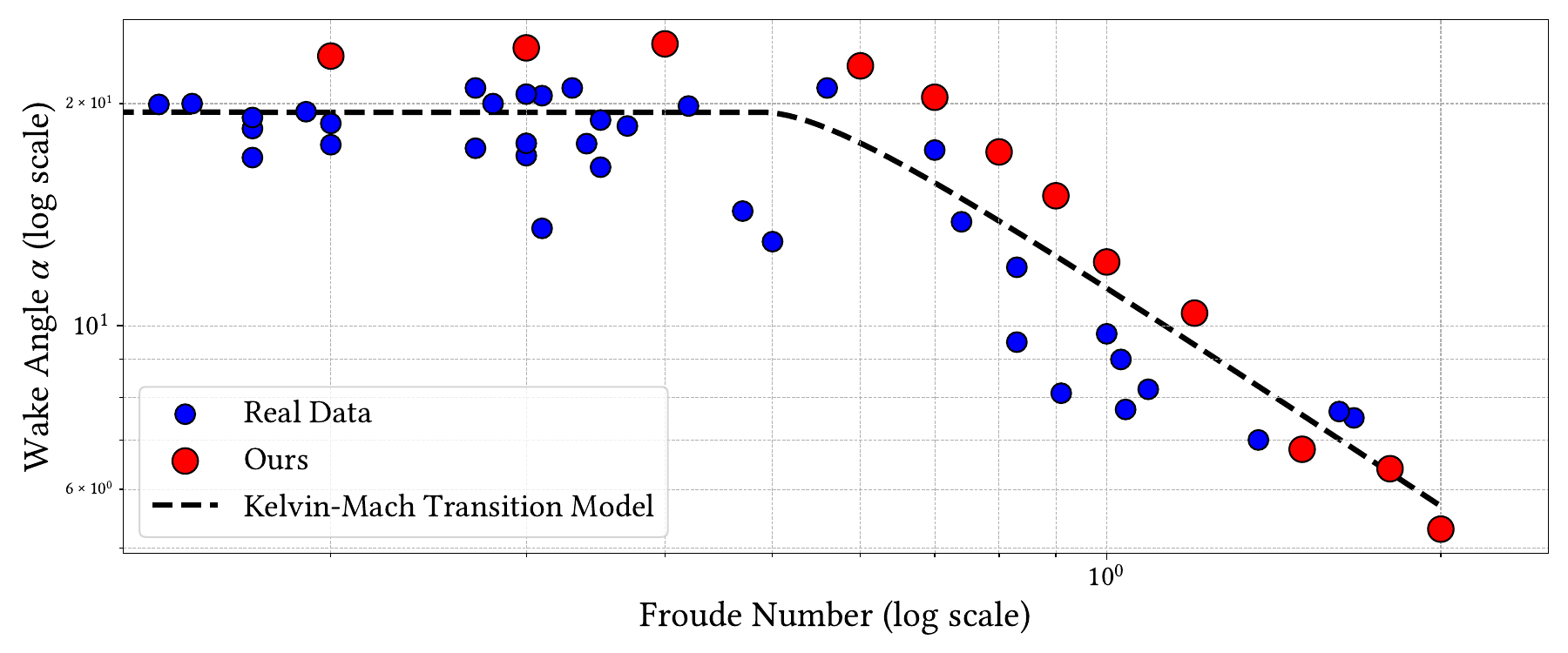}
    \caption{Wake angle vs. Froude Number ($Fr=U/\sqrt{gL}$) compared with the Kelvin--Mach Transition Model \cite{rabaud2013ship}. \rev{All runs use $\varepsilon=1$.}}
    \label{fig:fr}
\end{figure}

\begin{figure*}[!t]
    \centering
    \includegraphics[width=1.0\textwidth]{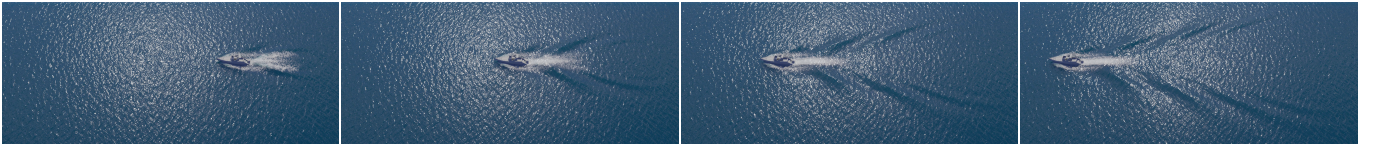}
    \caption{A \SI{4.8}{\meter}-long boat cruising at \SI{4}{m/s} across a \SI{72}{\meter}$\times$\SI{3}{\meter}$\times$\SI{36}{\meter} water domain. The boat produces a clean feather-like Kelvin wake \rev{\cite{rabaud2013ship}} whose long dispersive tail is carried far beyond the 3D fluid box by the 2D wave solver without visible discontinuity at the 2D--3D interface. \rev{The wave model uses $\varepsilon=1$.}}
    \label{fig:singleboat}
\end{figure*}

We simulate a \SI{4.8}{\meter}-long boat moving at \SI{4}{m/s} across a \SI{72}{\meter}$\times$\SI{3}{\meter}$\times$\SI{36}{\meter} water domain; \autoref{fig:singleboat} shows a rendering of this scene produced by our method with whitewater, where the feather-like Kelvin wake \rev{(``feather-like'' follows the terminology of \citet{rabaud2013ship})} is carried continuously across the 2D--3D interface. To assess our coupling scheme, we benchmark it against six alternatives under identical conditions, as shown in \autoref{fig:single_boat}: (a) \citet{chentanez2015coupling}, (b) the Airy DK wave model \cite{canabal2016dispersion} coupled via the FAB scheme \cite{stomakhin2017fluxed}, (c) our method with velocity relaxation, (d) our wave model combined with the coupling scheme of \citet{chentanez2015coupling}, (e) our full method, (f) a 3D NB-FLIP reference, and (g) a pure 3D Eulerian level-set reference. Method (a) produces a qualitatively wrong wake, as the SWE cannot resolve the dispersive Kelvin pattern. Methods (b), (c), and (d) reproduce the feather-like shape but exhibit clear spurious reflections behind the stern (highlighted by the red boxes in \autoref{fig:single_boat}). Our full method (e) \revv{correctly }reproduces the feather-like Kelvin wake with minimal spurious reflection at the 2D--3D interface. The NB-FLIP reference (f) closely matches our result, but its trailing wave at the far right is noticeably damped by the numerical dissipation of the 3D solver, while our coupled scheme preserves the wave because the 2D wave solver carries it without dissipation. The pure Eulerian level-set simulation (g) is included as an independent ground-truth reference. As reported in \autoref{statistics}, our method also runs more than $4\times$ faster than the pure GPU NB-FLIP simulation on this scene (\SI{85.0}{\milli\second} vs.\ \SI{363}{\milli\second} per substep). \rev{We note that the feather-like appearance is a qualitative descriptor; the wake-angle study below provides the corresponding quantitative validation of the simulated wakes (\autoref{fig:fr}).}

\rev{Specifically}, according to \cite{rabaud2013ship}, real-world ship wakes transition between the Kelvin angle and the Mach angle as the ship speed varies. We replicate this transition by simulating ship wakes at different speeds using the same experimental setting above. The resulting wake angles, illustrated in \autoref{fig:fr}, match the real-world data from \cite{rabaud2013ship} and correctly capture the Kelvin--Mach transition.

\subsection{Large-Scale 3D Demonstrations}

\paragraph{Submarine}

As shown in \autoref{fig:submarine}, to demonstrate the capability of our coupling scheme in handling complex fluid--solid interactions, we simulate a submarine submerging below the sea surface and gradually rising up. The simulation domain spans \SI{80}{\meter}$\times$\SI{40}{\meter} for the sea surface, with an embedded 3D region of \SI{24}{\meter}$\times$\SI{6}{\meter}$\times$\SI{12}{\meter} that includes a \SI{10}{\meter}-long submarine moving at \SI{4.3}{m/s}. As the submarine surfaces, it leaves a visible wake on the 2D ocean surface, while detailed 3D water motion is captured around the hull. To suppress nonlinear instabilities triggered by the rapidly moving hull, we cap the nonlinear amplitude parameter (see \autoref{sec:wave}) at $\varepsilon=0.3$.


\paragraph{Seaplane}

As shown in \autoref{fig:seaplane}, we simulate a seaplane landing on the ocean. The simulation domain spans \SI{160}{\meter}$\times$\SI{40}{\meter} for the sea surface, with an embedded 3D region of \SI{36}{\meter}$\times$\SI{6}{\meter}$\times$\SI{12}{\meter} that includes a seaplane with a \SI{10}{\meter} wingspan, pitched \SI{9}{\degree} nose-up and approaching at \SI{8}{m/s}. Upon touchdown, the aircraft skims the surface at \SI{10}{m/s}, leaving a long trailing wake that propagates \revv{seamlessly}\rev{smoothly} across the 2D--3D interface.


\paragraph{Battleship in Heavy Waves}

To further demonstrate the robustness of our coupling scheme, we simulate heavy ocean waves in a \SI{120}{\meter}$\times$\SI{60}{\meter} 2D domain and a \SI{24}{\meter}$\times$\SI{12}{\meter}$\times$\SI{12}{\meter} 3D fluid domain. In this scenario, the battleship is simulated using weak two-way solid--fluid coupling. The results in \autoref{fig:battleship} show that our simulation remains stable even with heavy waves, and the 2D and 3D \rev{simulations} remain \rev{visually consistent}. Because the background spectrum drives the surface close to the breaking threshold, we further restrict the nonlinear amplitude parameter (see \autoref{sec:wave}) to $\varepsilon=0.2$ to keep the wave evolution stable. \rev{This scene is our most demanding stress test, and the coupling is not entirely artifact-free: in the accompanying animation, wavefronts reflected from the rectangular boundary of the 3D box can be seen propagating outward into the 2D domain. Because the wave model is restricted to $\varepsilon=0.2$ for stability, it cannot fully absorb the strongly nonlinear surface content generated by the 3D solver under near-breaking forcing; the unabsorbed content instead radiates outward from the box boundary. \autoref{sec:implementation} discusses practical mitigation strategies at render time, i.e., strengthening the ocean-spectrum displacement layer and emitting foam and whitewater around the box footprint.}

\paragraph{Two Chasing Boats}

As shown in \autoref{fig:two_boat}, two boats with two independent 3D simulation domains are simulated on an ocean surface sized \SI{180}{\meter}$\times$\SI{90}{\meter}. The leading boat follows a curved path with one-way solid--fluid coupling. The trailing boat employs two-way solid--fluid coupling, responding dynamically to both the ambient waves and the wake generated by the leading boat.

\paragraph{Pond}

\begin{figure*}[!htbp]
    \centering
    \includegraphics[width=1.0\textwidth]{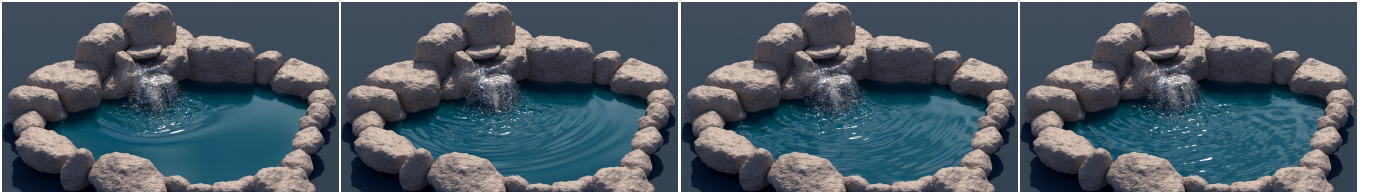}
    \caption{Water falling into a shallow pond with irregular boundaries. \rev{The wave model uses $\varepsilon=0.2$.}}
    \label{fig:pool}
\end{figure*}

We simulate water falling into a shallow pond with a depth of \SI{0.3}{\meter} and a 2D domain spanning \SI{2}{\meter}$\times$\SI{2}{\meter}, with the embedded 3D fluid box covering \SI{0.9}{\meter}$\times$\SI{0.45}{\meter}$\times$\SI{0.9}{\meter}. The pond outline and the rocks scattered inside it are encoded as an irregular solid mask on the 2D wave grid, while a separate rock mesh is treated as a rigid obstacle within the 3D fluid box. \autoref{fig:pool} illustrates falling flow that generates splashes and ripples that propagate across the surface and interact with the irregular solid boundaries. Before every spectral wave step, the solver fills the masked cells by iteratively averaging the height $\eta$ and vertical velocity $v$ from their fluid-side neighbors, producing a discrete zero-gradient extension of $(\eta,v)$ into the solid region. The subsequent FFT advance then sees this extension as replicated padding at the irregular boundary, so waves reflect off the pond walls and the rocks inside it at the cost of a few sweep iterations. Because the impact of the falling stream produces sharp local features that can excite high-wavenumber nonlinear instabilities, we cap the nonlinear amplitude parameter at $\varepsilon=0.2$ to keep the wave evolution stable.

%% file: conclusion_limitation.tex
\section{Conclusions and Limitations}

\paragraph{Conclusions}
We have presented a two-way coupled water solver that pairs a \revv{fully} nonlinear and \revv{Hamiltonian, }dispersive \revv{spectrally structured} 2D wave model with a localized 3D Navier--Stokes solver, \rev{addressing} a long-standing structural \rev{mismatch} in graphics where the 2D side of every prior coupling pipeline was either linear (Airy) or non-dispersive (SWE). \rev{Our solver makes the opposite trade-off to prior 2D wave models: whereas they buy speed and robustness by discarding nonlinearity, and remain the better choice when the surface stays gentle and nearly linear, we accept a modestly more expensive and less stable wave step in exchange for greater nonlinear accuracy and coupling fidelity that the strongly nonlinear scenes motivating a 2D--3D decomposition demand.} The 2D model evolves the canonical Zakharov pair $(\eta,\psi)$ under the exact water-wave Hamiltonian, advancing the Dirichlet--Neumann operator through a homogeneous Craig--Sulem expansion so that every nonlinear correction reduces to FFTs and pointwise products on a regular grid at $\mathcal{O}(N^2\log N)$ cost. Because the wave side and the 3D bulk solver carry the same physical variable, the surface velocity potential, the coupling becomes natural: both directions of state transfer are realized as Dirichlet--Neumann maps, and a Newtonian relaxation source integrated through its exact exponential decay returns the 3D-side surface state to the canonical wave variables without damping outgoing waves. \rev{The Hamiltonian structure of the wave model thus pays off chiefly in the coupling: the canonical pair $(\eta,\psi)$ serves as a complete interface state that both solvers can read and write consistently.} A scalar amplitude parameter $\varepsilon\in[0,1]$ continuously interpolates between the linear Airy limit and the \revv{fully} nonlinear Zakharov system, giving the user a single dial to trade nonlinear fidelity against stability. We validate our method on a wide range of experiments, including standalone 2D wave-propagation benchmarks, controlled 2D--3D coupling tests, and a suite of large-scale production-style scenarios such as a surfacing submarine, a seaplane landing, a battleship in heavy seas, two chasing boats, and a shallow pond with irregular boundaries. Across all these cases, our coupled pipeline produces accurate\revv{, artifact-free} results \rev{with minimal coupling artifacts} at competitive runtime, \rev{suggesting that the method is practical and applicable across a broad range of scenarios within its intended nonlinear regime}.

\paragraph{Limitations}
Our solver has \rev{several} limitations. First, the fully nonlinear setting ($\varepsilon=1$) is not unconditionally stable: when strong 3D forcing pushes the surface close to the breaking threshold, the higher-order DNO terms can excite numerical instabilities, and we have to reduce $\varepsilon$ to maintain a stable simulation. In our experiments, the submarine uses $\varepsilon=0.3$, while the battleship in heavy waves and the pond both use $\varepsilon=0.2$. While our $\varepsilon$ sweeps (\autoref{fig:wave_prop_err},~\autoref{fig:eps_sweep_err}) confirm that even a small $\varepsilon$ already removes a large fraction of the linear-model error, the user is still left to tune $\varepsilon$ per scene rather than always running at full nonlinearity. Second, the truncated DNO introduces a computational overhead that grows with the truncation order: as reported in \autoref{tab:wave_prop_time_comp}, Order 3 is roughly $2\times$ slower than Order 2 per step, and each additional order adds a comparable cost. \rev{More broadly, even the Order-2 wave step is roughly $5\times$ more expensive than linear-wave solvers such as SWE and Airy DK (\autoref{tab:wave_prop_time_comp}); in our coupled pipeline this overhead is minor in absolute terms, since the 3D solver dominates the total cost in most of our scenes (\autoref{statistics}), but for standalone ocean animation where a linear model suffices, our solver is the more expensive choice.} In practice, we find that Order 2 is already sufficient for the vast majority of our scenes, with Order 3 providing only marginal accuracy gains in the steepest regimes (\autoref{fig:eps_sweep_err}). Third, our 2D wave model cannot represent breaking waves: once the surface overturns, the single-valued height field $\eta(\bm{x},t)$ ceases to be a valid representation. \rev{We also note that the wave solver advances the entire 2D grid at every substep, so unlike Tessendorf-style spectral oceans, the simulated wave domain cannot be extended indefinitely. Finally, our coupling does not fully eliminate interface artifacts in the most demanding scenarios, such as the battleship scene, and we currently rely on render-time masking to visually mitigate these residual artifacts (\autoref{sec:implementation}).}

\paragraph{Future Work}
Each limitation points to a distinct direction. To address the conditional stability of the fully nonlinear model, we would like to investigate more stable nonlinear time-integration schemes for the Zakharov system, such as Hamiltonian-preserving integrators, spectrally consistent dealiasing tailored to the higher-order DNO terms, or implicit treatments of the leading nonlinear corrections, with the goal of pushing the stable operating point of $\varepsilon$ closer to $1$ even in strongly forced regimes. To reduce the per-step cost of the nonlinear correction, we will explore cheaper approximations of the higher-order DNO terms, for example mixed-precision FFT pipelines, learned surrogates for $G_j(\eta)$ at $j\ge 2$, or adaptive-order schemes that locally raise the truncation order only where the surface is steep. Finally, to extend the method to breaking and overturning waves, we plan to swap out the single-valued height-field representation in regions where breaking is detected for a richer surface representation, such as a locally multi-valued surface, a particle-augmented free surface in the spirit of NB-FLIP, or a hybrid Hamiltonian/Lagrangian description that retains the spectral efficiency of our solver in the smooth bulk while gracefully degrading to a particle representation through the breaking event.

\begin{acks}
We sincerely thank the reviewers for their valuable feedback. Georgia Tech authors acknowledge NSF CAREER \#2420319, IIS \#2433307, OISE \#2433313, IIS \#2433322, ECCS \#2318814, and the NVIDIA Academic Grant for funding support. We credit the Houdini education license for video animations.
\end{acks}